\documentclass[preprint]{aastex}
\singlespace

\usepackage{natbib}

\slugcomment{to appear in Astronomical Journal, 2002}

\newcommand{\OO}[1]{{\cal O}(c^{-#1})}
\newcommand{\vecg}[1]{\mbox{\boldmath$#1$}}
\newcommand{\ve}[1]{\vecg{#1}}
\newcommand{\muas}[0]{\hbox{\rm $\mu$as}}

\def\TCG{{\hbox{\rm TCG}}}
\def\TCB{{\hbox{\rm TCB}}}
\def\TDB{{\hbox{\rm TDB}}}
\def\TT{{\hbox{\rm TT}}}

\shorttitle{A Practical Relativistic Model for Microarcsecond Astrometry in Space}
\shortauthors{S.A. Klioner}

\begin{document}

\title{A Practical Relativistic Model for Microarcsecond Astrometry
in Space}

\author{Sergei A. Klioner}
\affil{Lohrmann Observatory, Dresden Technical University, \\
Mommsenstr. 13, 01062 Dresden, Germany}
\email{klioner@rcs.urz.tu-dresden.de}

\begin{abstract}
This paper is devoted to a practical model for relativistic reduction
of positional observations with an accuracy of 1 \muas\ which is
expected to be attained in the future space astrometry missions. All
relativistic effects which are caused by the gravitational field of the
Solar system and which are of practical importance for this accuracy
level are thoroughly calculated and discussed. The model includes
relativistic modeling of the motion of observer, modeling of
relativistic aberration and gravitational light deflection as well as a
relativistic treatment of parallax and proper motion suitable for the
accuracy of 1 \muas. The model is formulated both for remote sources
(stars, quasars, etc.) and for Solar system objects (asteroids, etc.).
The suggested model is formulated within the framework of Parametrized
Post-Newtonian Formalism with parameters $\beta$ and $\gamma$. However,
for general relativity ($\beta=\gamma=1$) the model is fully compatible
with the IAU Resolutions (2000) on relativity in celestial mechanics,
astrometry and metrology. The model is presented in a form suitable for
implementation in a software system for data processing or simulation.
The changes which should be applied to the model to attain the accuracy
of 0.1 \muas\ are reviewed. Potentially important relativistic effects
caused by additional gravitational fields which are generated outside
of the Solar system are also briefly discussed.
\end{abstract}


\section{Introduction}

Within the next decade the accuracy of space-based astrometric
positional observations is expected to attain a level of 1
microarcsecond (\muas). The problem of relativistic modeling of
positional observations with a microarcsecond accuracy has become quite
practical in the recent years when a number of astrometric space
projects have been approved by NASA, ESA and other boards and selected
for launch in the next several years (GAIA
\citep{GAIA:2000,Perryman:et:al:2001,Bienayme:Turon:2002}, SIM
\citep{SIM:1998}, etc.). Following this practical trend the present
paper describes a practical relativistic model of space-based
positional observations which is valid at a level of 1 \muas\ and which
can be readily implemented in the corresponding software.



Relativistic effects in positional observations have been studied by
many authors in many different aspects. Needless to say that the
gravitational light deflection in the gravitational field of the Sun
being one of the most important constituents of the relativistic model
of positional observations was one of the first experimental tests of
general relativity. Already in the time of Hipparcos it was realized
that relativity must play an important role in the formulation of
the transformation between the observed positions of a star and what should
be included in the resulting catalogue \citep{Walter:et:al:1986}.
However, it was a relatively uncomplicated task to formulate such a
transformation for Hipparcos which was aimed at the final accuracy of 1 mas. A
model of positional observations suitable for an accuracy of 1 \muas\
is much more intricate. It is clear since in this case typical
relativistic effects exceed the required accuracy by several
orders of magnitude. The whole concept of the modeling should be
re-formulated in the framework of general relativity. Besides that at
such a high level of accuracy many additional, more subtle relativistic
effects should be taken into account.


The first complete general-relativistic model of positional
observations at the microarcsecond level of accuracy was formulated by
\citet{Klioner:Kopeikin:1992}. That work was stimulated by
the early project for microarcsecond astrometry in space POINTS
\citep{Reasenberg:et:al:1988}. The model was formulated in a rather
general form and was primarily intended for a satellite on a geocentric
orbit (geostationary or lower). The model described in this paper is
based on the same general post-Newtonian approximation scheme used in
the model of \citet{Klioner:Kopeikin:1992}, but differs from the latter
in several important aspects: (1) the Geocentric Celestial Reference
System (see below) is used only as an intermediate reference system to
model Earth-based observations of the satellite itself (orbit
determination) and not to process the astrometric observations produced
by the satellite; (2) the model contains a refined treatment of several
relativistic effects (primarily aberration and some subtle effects in
the gravitational light deflection); (3) the model is formulated in the
framework of the so-called Parametrized Post-Newtonian (PPN) formalism
\citep{Will:1993} which makes it possible to use the positional
observations to test general theory of relativity; (4) the present
model is suitable both for remote sources located outside of the Solar
system and for Solar system objects, the coupling between the finite
distance to the object and the gravitational light deflection being
properly treated; (5) the model is supplemented with a set of simple
formulas allowing one to judge if a particular gravitational light
deflection effect due to a particular gravitating body must be taken
into account or can be neglected for a particular mutual disposition of
the source, the observer and the gravitating body; (6) the model is
optimized and simplified as much as possible for the goal accuracy of 1
\muas, which makes it straightforward to implement the model in the
corresponding software.


Since the practical interest in microarcsecond astrometry was revived
in the last years, a number of numerical simulations of astrometric
missions has appeared.
\citet{deFelice:Lattanzi:Vecchiato:Bernacca:1998,
deFelice:Vecchiato:Bucciarelli:Lattanzi:Crosta:2000,
deFelice:Bucciarelli:Lattanzi:Vecchiato:2001} have used a simplified
relativistic model (Schwarzschild field of the Sun) in an end-to-end
simulation of the GAIA mission to demonstrate its capability and
investigate various statistical properties of the solution in different
cases. Simulations based on more realistic models are in progress
\citep{Kopeikin:Shuygina:Vasiliev:Yagudina:Yagudin:2000,
deFelice:Bucciarelli:Lattanzi:Vecchiato:2001}. The model represented in
this paper can be used to facilitate further investigations of this
kind.

It is clear that the model given below is not all what is needed for
practical processing of the data produced by an astrometric satellite such
as GAIA (or for corresponding data simulations). The full model of
observables should contain not only the idealized relativistic part
given below, but also a detailed instrumental model describing
observables from the ``technical'' point of view. The ``technical''
model heavily depends on the particular design of the satellite and
should relate theoretical ``observable direction'' toward a source to
the technical observational data provided by the particular satellite
design. This ``technical'' model may contain parameters of the
satellite's hardware, various calibration parameters, parameters
describing the attitude of the satellite, etc. Besides that, the
mathematical and statistical details of the data processing of the
future astrometric missions represent an important research topic in
itself. All these aspects of practical data processing are beyond the
scope of the present paper.


Let us summarize the most important notations used throughout the
paper:

\begin{itemize}

\item $G$ is the Newtonian constant of gravitation;

\item $c$ is the velocity of light;

\item $\beta$ and $\gamma$ are the parameters of the Parametrized Post-Newtonian
(PPN) formalism which characterize possible deviation of the physical
reality from general relativity theory ($\beta=\gamma=1$ in general
relativity);

\item the lower case latin indices $i$, $j$, $k$, \dots take values 1,
2, 3;

\item the lower case latin indices are lowered and raised by means of
the unit matrix $\delta_{ij}=\delta^{ij}={\rm diag}(1,1,1)$.
Therefore, the disposition of such indices plays no
role: $a^i=a_i$;

\item repeated indices imply the Einsteinian summation irrespective of
their positions (e.g. $a^i\,b^i=a^1\,b^1+a^2\,b^2+a^3\,b^3$);

\item a dot over any quantity designates the total derivative with
respect to the coordinate time of the corresponding reference system:
e.g. $\dot a=\displaystyle{da\over dt}$;

\item the 3-dimensional coordinate quantities (``3-vectors'') referred to
the spatial axes of the corresponding reference system are set in
boldface: $\ve{a}=a^i$;

\item the absolute value (Euclidean norm) of a ``3-vector'' $\ve{a}$ is
denoted as $|\ve{a}|$ and can be computed as
$|\ve{a}|=(a^1\,a^1+a^2\,a^2+a^3\,a^3)^{1/2}$;

\item the scalar product of any two ``3-vectors'' $\ve{a}$ and $\ve{b}$
with respect to the Euclidean metric $\delta_{ij}$ is denoted by
$\ve{a}\,\cdot\,\ve{b}$ and can be computed as
$\ve{a}\,\cdot\,\ve{b}=\delta_{ij}\,a^i\,b^j=a^i\,b^i$;

\item the vector product of any two ``3-vectors'' $\ve{a}$ and $\ve{b}$
is designated by $\ve{a}\times\ve{b}$ and can be computed as
$\left(\ve{a}\times\ve{b}\right)^i=\varepsilon_{ijk}\,a^j\,b^k$, where
$\varepsilon_{ijk}=(i-j)(j-k)(k-i)/2$ is the fully antisymmetric
Levi-Civita symbol;

\item the capital italic subscripts $A,B,C,\dots = 1 \dots N$ refer to
the gravitating bodies of the Solar system; as a special case
subscript $E$ designates quantities related to the Earth;

\item the subscript '$o$' denotes quantities related to the observer
(satellite): e.g. $\ve{x}_o$ denotes the position of the observer and
$t_o$ is the coordinate time of observation with respect to the
Barycentric Celestial Reference System (BCRS) of the IAU (see below);

\item the subscript '$p$' denotes quantities related to the light ray
(photon): e.g. $\ve{x}_p(t)$ denotes the BCRS position of the light ray
at some moment of time $t$;

\item the subscript '$s$' denotes quantities related to the source:
e.g. $\ve{x}_s$ denotes the BCRS position of the source;

\item the subscript '$e$' denotes quantities related to the moment of
emission of the light ray by the source: e.g. $t_e$ is the BCRS
coordinate time of emission of the signal by the source;

\end{itemize}


Section \ref{Section-general-scheme} is devoted to a general scheme of
relativistic modeling of astronomical observations of any kind in the
framework of general relativity or PPN formalism. The overall structure
of the specific modeling scheme for positional observations made from a
space station is described in Section
\ref{Section-general-scheme-positional}. Modeling of the motion of the
observer (satellite) and that of its proper time are discussed in Section
\ref{Section-motion}. Section \ref{Section-aberration} deals with the
relativistic description of aberration. Gravitational light deflection
is discussed in Section \ref{Section-gravitational-deflection}.
Parallax and proper motion are analyzed in Sections
\ref{Section-parallax} and \ref{Section-proper-motion}, respectively.
Section \ref{Section-summary} summarizes the suggested relativistic
model. Section \ref{Section-0.1-muas} contains a short discussion of
the changes which should be applied to the model to attain the accuracy
of 0.1~\muas. In Section \ref{Section-beyond-the-model} several known
relativistic effects beyond the given model are described.

\section{General scheme of relativistic modeling of astronomical
observations}
\label{Section-general-scheme}

\begin{figure}
\unitlength=1.0mm
\def\betweenlines{0mm}
\linethickness{0.5pt}
\font\rs=cmss10 scaled \magstephalf
\font\rrs=cmss10 scaled \magstep1
\begin{center}
\begin{picture}(155.00,180.00)(5.00,-15.00)
\baselineskip=0pt
\parskip=0pt
\put(40.00,143.00){\framebox(70.00,15.00)[cc]{\parbox{60mm}{\begin{center}\rs METRIC THEORY\\[\betweenlines] OF GRAVITY\vbox to 3mm{} \end{center}}}}
\put(115.00,150.00){\parbox{80mm}{\rrs {General relativity or}\\[\betweenlines] {PPN formalism}}}
\put(40.00,115.00){\framebox(70.00,17.00)[cc]{\parbox{60mm}{\begin{center}\rs A SET OF ASTRONOMICAL\\[\betweenlines] REFERENCE SYSTEMS\end{center}}}}
\put(102.00,76.00){\framebox(45.00,26.00)[cc]{\parbox{40mm}{\begin{center}\rs RELATIVISTIC\\[\betweenlines] DESCRIPTION\\[\betweenlines] OF THE PROCESS\\[\betweenlines] OF OBSERVATION\end{center}}}}
\put(2.00,76.00){\framebox(45.00,26.00)[cc]{\parbox{40mm}{\begin{center}\rs RELATIVISTIC\\[\betweenlines] EQUATIONS\\[\betweenlines] OF MOTION\end{center}}}}
\put(52.00,76.00){\framebox(45.00,26.00)[cc]{\parbox{40mm}{\begin{center}\rs EQUATIONS OF\\[\betweenlines] SIGNAL\\[\betweenlines] PROPAGATION\end{center}}}}
\put(75.00,113.00){\vector(0,-1){9.00}}
\put(52.00,38.00){\framebox(45.00,25.00)[cc]{\parbox{40mm}{\begin{center}\rs RELATIVISTIC\\[\betweenlines] MODELS\\[\betweenlines] OF OBSERVABLES\end{center}}}}
\put(75.00,74.00){\vector(0,-1){9.00}}
\put(25.00,74.00){\vector(4,-1){36.00}}
\put(125.00,74.00){\vector(-4,-1){34.00}}
\put(61.00,113.00){\vector(-4,-1){36.00}}
\put(91.00,113.00){\vector(4,-1){34.00}}
\put(75.00,141.5){\vector(0,-1){8.00}}
\put(75.00,36.00){\vector(0,-1){9.00}}
\put(113.00,0.00){\framebox(40.00,25.00)[cc]{\parbox{40mm}{\begin{center}\rs OBSERVATIONAL\\[\betweenlines] DATA\end{center}}}}
\put(50.00,0.00){\framebox(50.00,25.00)[cc]{\parbox{40mm}{\begin{center}\rs ASTRONOMICAL\\[\betweenlines] REFERENCE\\[\betweenlines] FRAMES\end{center}}}}
\put(10.00,30.00){\parbox{40mm}{\rrs {Coordinate-}\\[\betweenlines]\hbox to 5mm{} {dependent}\\[\betweenlines] {parameters}}}
\put(40.00,33.00){{\vector(+1,+3){5.00}}}
\put(40.00,29.00){{\vector(+1,-3){5.00}}}
\put(111.00,12.00){\vector(-1,0){9.00}}
\end{picture}
\end{center}
\caption{General principles of relativistic modeling of astronomical observations
(see text for further explanations).
\label{Figure-general-principles}
}
\end{figure}
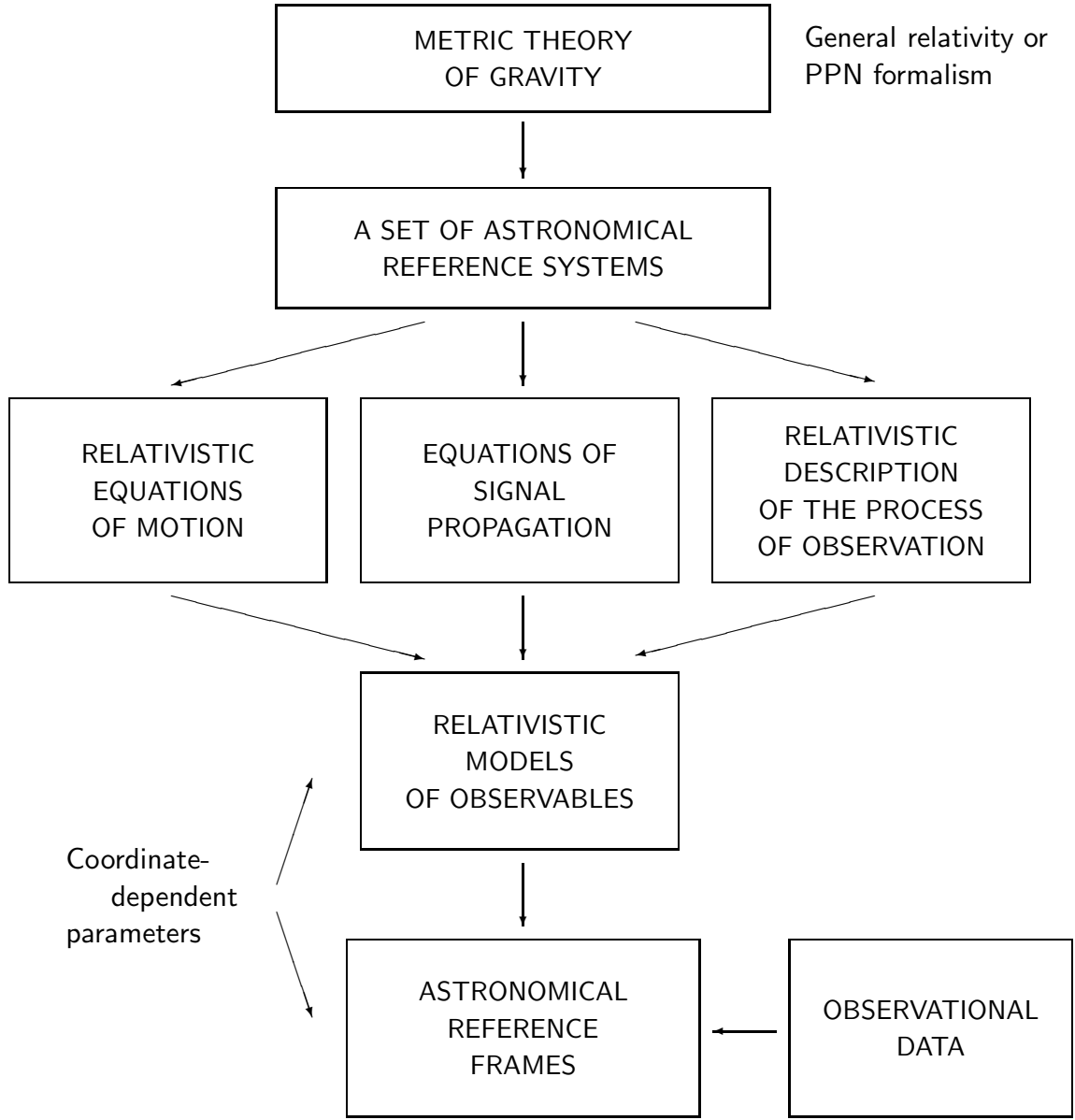

\begin{figure}
\epsscale{0.8}
\plotone{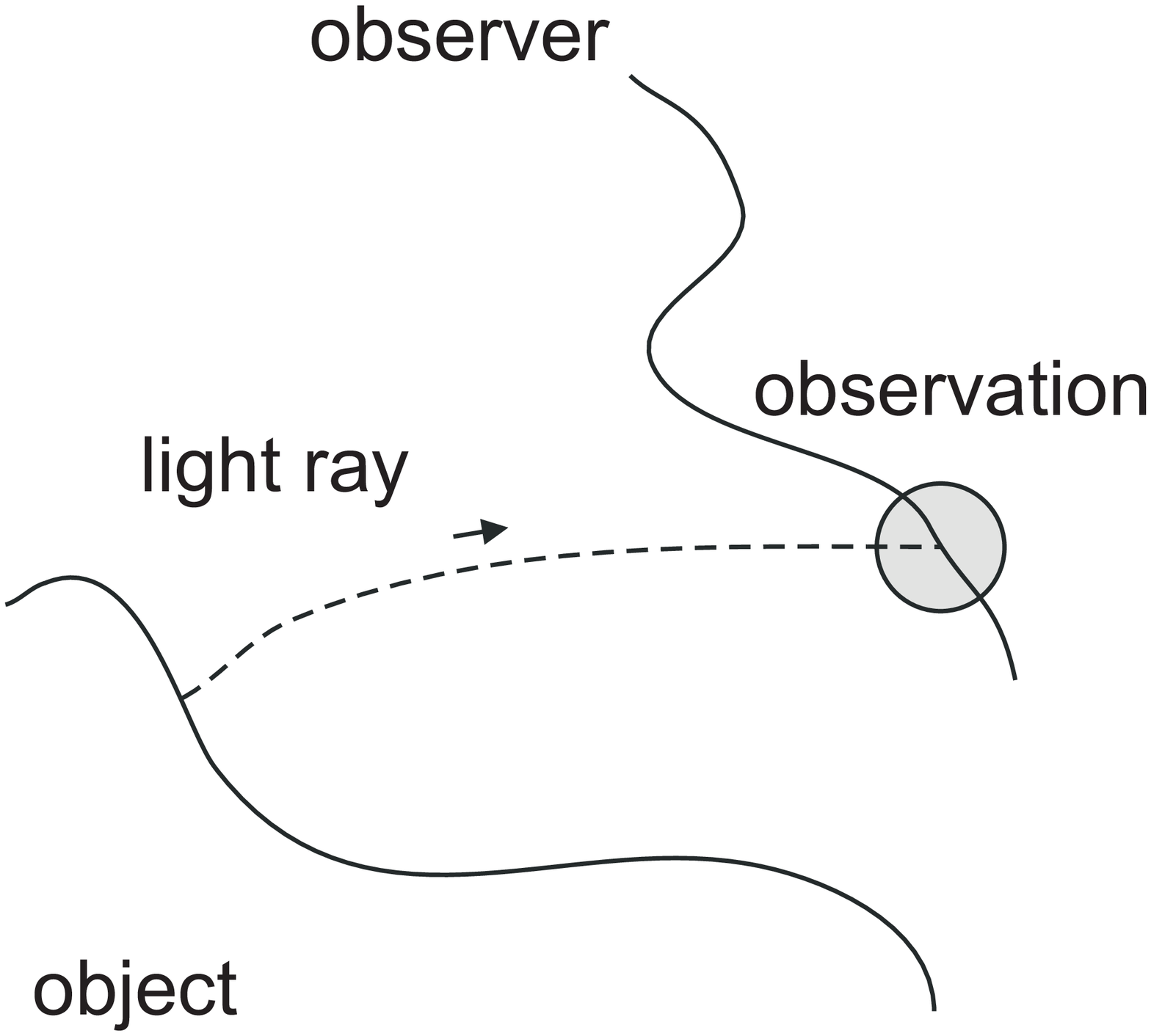}
\caption{Four constituents of an astronomical event: 1) motion of
the observed object, 2) motion of the observer, 3) propagation of an
electromagnetic signal from the observed object to the observer, 4)
the process of observation.
\label{Figure-observation}
}
\end{figure}

Let us first outline general principles of relativistic modeling of
astronomical observations. General scheme is represented on Fig.
\ref{Figure-general-principles}. Starting from general theory of
relativity, any other metric theory of gravity or PPN formalism one
should define at least one relativistic 4-dimensional reference system
covering the region of space-time where all the processes constituting
particular kind of astronomical observations are located. Typical
astronomical observation depicted on Fig. \ref{Figure-observation}
consists of four constituents: motion of an observer, motion of an
observed object, light progagation and the process of observation. Each
of these four constituents should be modeled in the relativistic
framework. The equations of motion of both the observed object and the
observer relative to the chosen reference system should be derived and
a method to solve these equations should be found. Typically the
equations of motion are second-order ordinary differential
equations and numerical integration with suitable initial or boundary
conditions can be used to solve them. Astrometric information on the
object can be read off the electromagnetic signals propagating from the
object to the observer. Therefore, the corresponding equations of light
propagation relative to the chosen reference system should be derived
and solved.  The equations of motion of the object and the observer and
the equations of light propagation enable one to compute positions and
velocities of the object, observer and the photon (light ray) with
respect to the particular reference system at a given moment of
its coordinate time, provided that the positions and velocities at some
initial epoch are known. However, these positions and velocities
obviously depend on the used reference system. On the other hand, the
results of observations cannot depend on the reference system used to
theoretically model the observations. Therefore, it is clear that one
more step of the modeling is needed: a relativistic description of the
process of observation. This part of the model allows one to compute a
coordinate-independent theoretical prediction of the observables
starting from the coordinate-dependent position and velocity of the
observer and, in some cases, the coordinate velocity of the
electromagnetic signal at the point of observation.

Mathematical techniques to derive the equations of motion of the
observed object and the observer, to formulate the equations of light
propagation and to find the description of the process of observation
in the relativistic framework are well known and will be discussed
below. These three parts can now be combined into relativistic models
of observables. The models give an expression for each observable under
consideration as a function of a set of parameters. These parameters
can then be fitted to observational data using some kind of
parameter estimation scheme (e.g. least squares or other estimators).
The sets of certain estimated parameters appearing in the relativistic
models of observables represent astronomical reference frames.

Note that {\sl reference system} is a purely mathematical construction
(a chart) giving ``names'' to space-time events. In contrast to this a
{\sl reference frame} is some materialization of a reference system. In
astronomy usually the materialization is realized in a form of a
catalogue (or ephemeris) containing positions of some celestial objects
relative to the reference system under consideration. Any reference
frame (a catalogue, an ephemeris, etc) is defined only through the
reference system(s) used to construct physical models of observations.

It is very important to understand at this point that the relativistic
models contain parameters which are defined only in the chosen
reference system(s) and are thus coordinate-dependent. A good example of
such coordinate-dependent parameters are the coordinates and velocities
of various objects (e.g. major planets or the satellite) at some
epoch. On the other hand, from the physical point of view any reference
system covering the region of space-time under consideration can be used
to describe physical phenomena within that region, and we are free to
choose the reference system to be used to model the observations.
However, reference systems, in which mathematical description of
physical laws is simpler than in others, are more convenient for
practical calculations. Therefore, one can use the freedom to choose
the reference system to make the parametrization as convenient and
reasonable as possible (e.g. one prefers the parameters to have
a simpler time dependence).

For modeling of physical phenomena localized in some sufficiently small
region of space (e.g. in the vicinity of a massless observer or a
gravitating body) one can construct a so-called local reference system
where the gravitational influence of the outer world is effaced as much
as possible in accordance with the Einstein equivalence principle. In
the local reference system of a material system the gravitational field
of the outer matter manifests itself in the form of tidal gravitational
potential. First, the Solar system as a whole can be considered as one
single body, and a reference system can be constructed where the
gravitational influence of the matter situated outside of the Solar
system can be described by a tidal potential. That tidal potential
(mainly due to the influence of the Galaxy) is utterly small and can be
then neglected for most purposes. Such a reference system is often
called barycentric reference system of the Solar system. It can be used
far beyond the Solar system and is suitable to describe the dynamics of
the Solar system (motion of planets and spacecraft relative to the
barycenter of the Solar system) as well as to model the influence of
the gravitational field of the Solar system on the light rays
propagating from remote sources to an observer. Second, the
corresponding local reference system can be constructed for the Earth.
Such a geocentric reference system is convenient to model the
geocentric motion of Earth satellites, the rotational motion of the
Earth itself, etc. Underlying theory and technical details concerning
local reference systems have been discussed in a series of papers by
Brumberg and Kopeikin \citep[see also
\citet{Klioner:Voinov:1993}]{Kopeikin:1988,Brumberg:Kopeikin:1989a,Brumberg:Kopeikin:1989b,Brumberg:1991}
and by \citet{Damour:Soffel:Xu:1991,Damour:Soffel:Xu:1992,Damour:Soffel:Xu:1993,Damour:Soffel:Xu:1994}.
\citet{IAU:2001} has recently recommended the use of a particular form of
the barycentric and local geocentric reference systems for modeling of
astronomical observations. These two standard relativistic reference
systems are called Barycentric Celestial Reference System (BCRS) and
Geocentric Celestial Reference System (GCRS). Coordinate time $t$ of
the BCRS is called Barycentric Coordinate Time (\TCB). Coordinate time
$T$ of the GCRS is called Geocentric Coordinate Time (\TCG). According
to the IAU notations, throughout this paper the spatial coordinates of
the BCRS will be designated as $\ve{x}$ while those of the GCRS as
$\ve{X}$.


Both Brumberg-Kopeikin and Damour-Soffel-Xu theories are based on
Einstein's general relativity theory. It is clear, however, that one
of the important goals of the future astrometric missions is to test
general relativity. The model presented in this paper will be given in
the framework of the PPN formalism \citep[see, e.g.][]{Will:1993}
including two main parameters $\beta$ and $\gamma$ which characterize
possible devitations of the physical reality from general relativity. Both
parameters are equal to unity in general relativity. The theory of
local reference systems with parameters $\beta$ and $\gamma$ has been
constructed by \citet{Klioner:Soffel:1998b,Klioner:Soffel:2000}. Both
that theory of local PPN reference systems and the model given in the
present paper are constructed in such a way that for $\beta=\gamma=1$
all the formulas coincide with those which can be derived directly from
the general-relativistic versions of the BCRS and GCRS recommended by
\citet{IAU:2001}.


\section{General structure of the relativistic model of positional
observation}
\label{Section-general-scheme-positional}

\begin{figure}
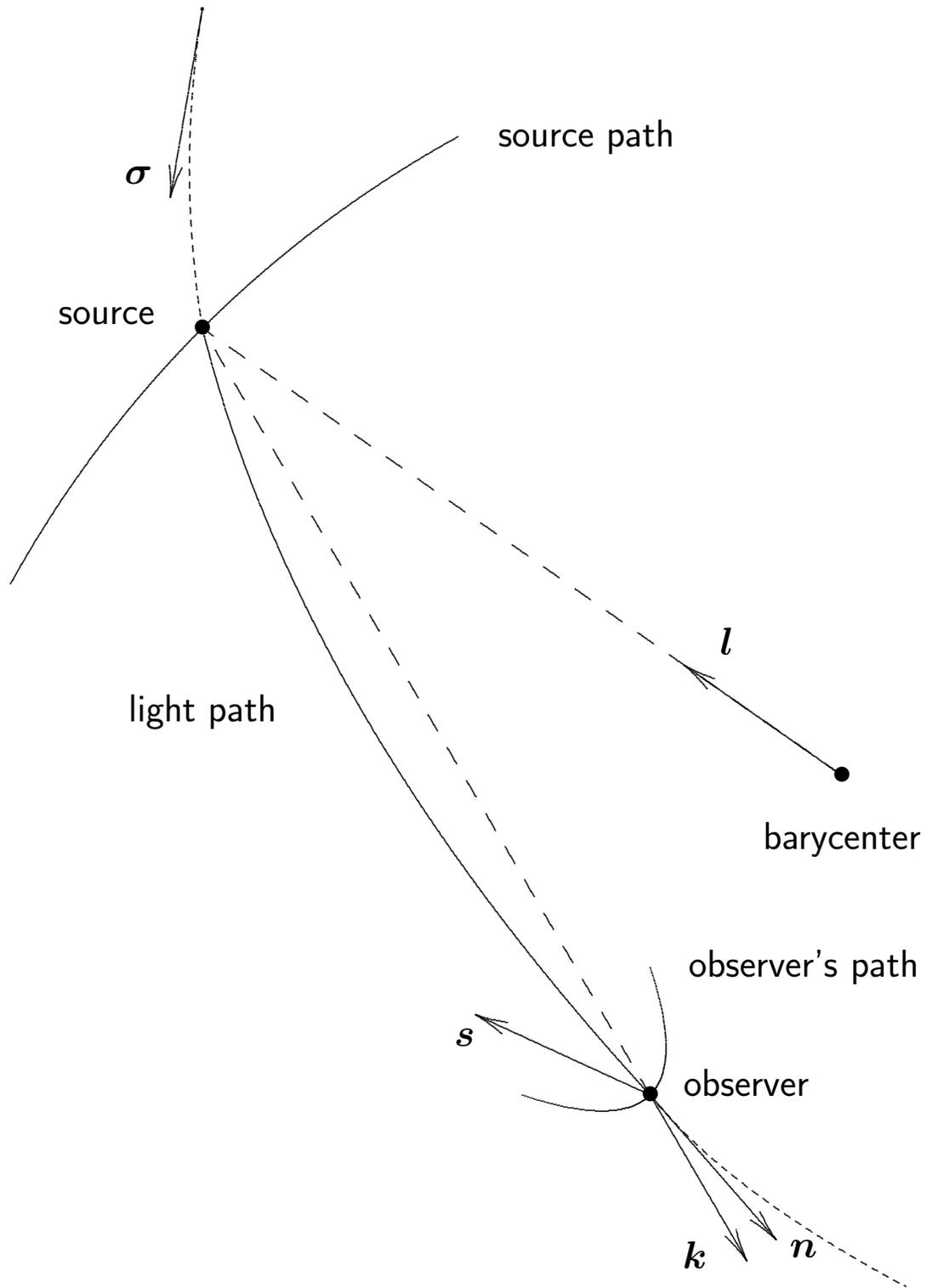

\font\fiverm=cmr5
\input prepicte.tex
\input pictex.tex
\input postpict.tex
\LARGE
\sf
\beginpicture
\setcoordinatesystem units <1mm, 1mm>
\setplotarea x from 0 to 200, y from 0 to 200
\put {$\bullet$} at 120 30
\put {$\bullet$} at 150 80
\put {$\bullet$} at 50  150
\arrow <5mm> [.15,.4] from 150 80 to 125.423 97.2039  
\put {$\ve{l}$} at 132 101
\setdashpattern <3mm, 3mm>
\setlinear
\plot 150 80  50 150 /
\plot 120 30  50 150 /
\setsolid
\setquadratic
\plot 20 110 50 150 90 180 /
\plot 100 30 120 30 120 50 /
\plot 120 30 75 90 50 150 /
\setdashpattern <1mm, 1mm>
\plot 50 150 48 175 50 200 /
\setsolid
\setlinear
\arrow <5mm> [.15,.4] from 50 200 to 45 170.42
\put {.} at  50.0 200
\put {$\vecg{\sigma}$} at 40 174
\put {source} at 35 152
\put {source path} at 110 180
\put {barycenter} at 150 70
\put {observer's path} at 144 50
\put {observer} at 135 32
\arrow <5mm> [.15,.4] from 120 30 to 92.745 42.537
\put {$\ve{s}$} at 91 39
\arrow <5mm> [.15,.4] from 120 30 to 135.1161308 4.08663298 
\put {$\ve{k}$} at  127 5
\arrow <5mm> [.15,.4] from 120 30 to 139.6853614 7.361834335 
\put {$\ve{n}$} at  144 6
\put {light path} at 50 90
\setdashpattern <1mm, 1mm>
\setquadratic
\plot 120 30 136 15 160 0 /
\endpicture
\caption{Five principal vectors used in the model: $\ve{s}$, $\ve{n}$,
$\vecg{\sigma}$, $\ve{k}$, $\ve{l}$. See text for further details.
\label{Figure-principal-vectors}
}
\end{figure}

The relativistic model of positional observations relates the observed
direction of the light coming from a source to the coordinates of that
source at the moment of emission. A set of the coordinate directions
toward the source for different moments of time can be then used to
obtain further parameters of the source describing its spatial position
and spatial motion with respect to the BCRS (parallax and proper motion
or barycentric orbital elements). It is convenient to divide the
conversion of the observed directions into the coordinate ones into
several steps. Let us introduce five vectors which will be used below:
$\ve{s}$ is the unit observed direction (the word ``unit'' means here
and below that the formally Euclidean scalar product
$\ve{s}\,\cdot\,\ve{s}= s^i\,s^i$\ is equal to unity), $\ve{n}$ is the
unit tangent vector to the light ray at the moment of observation,
$\vecg{\sigma}$ is the unit tangent vector to the light ray at
$t=-\infty$, $\ve{k}$ is the unit coordinate vector from the source to
the observer, $\ve{l}$ is the unit vector from the barycenter of the
Solar system to the source (see Fig. \ref{Figure-principal-vectors}).
Note that the last four vectors are defined formally in the coordinate
space of the BCRS and should not be interpreted as ``Euclidean''
vectors in some ``Newtonian physical space''. For the same physical
situation these vectors are different if different reference systems
are used instead of the BCRS. The word ``vector'' is used here to refer
to a set of three real numbers defined in the coordinate space of the
BCRS, rather than to a geometric object. A slightly different meaning
of $\ve{s}$ is discussed in Section~\ref{Section-aberration} below.

Apart from the modeling of the motion of the observer which will be
considered in the next Section, the model consists in subsequent
transformations of these five vectors. Namely, the following effects
should be subsequently considered:

\begin{itemize}

\item aberration (the effects related to the motion of the observer
with respect to the barycenter of the Solar system): this converts the
observed direction to the source $\ve{s}$ into the unit BCRS coordinate
direction of the light ray $\ve{n}$ at the point of observation;

\item gravitational light deflection for the source at infinity:  this
step converts $\ve{n}$ into the unit direction of propagation
$\vecg{\sigma}$ of the light ray infinitely far from the Solar system
for $t\to-\infty$;

\item coupling of the finite distance to the source and the
gravitational light deflection in the gravitational field of the Solar
system:  this step converts $\vecg{\sigma}$ into the unit BCRS vector
$\ve{k}$ going from the source to the observer (note that as discussed
below this step should be combined with the previous one for the
sources situated within the Solar system);

\item parallax: this step converts $\ve{k}$ into the unit vector $\ve{l}$
going from the barycenter of the Solar system to the source;

\item proper motion: this step provides a reasonable parametrization of
the time dependence of $\ve{l}$ caused by the motion of the source with
respect to the BCRS.

\end{itemize}

\noindent
All these steps will be specified in detail in the following Sections.
However, let us first clarify the question of time scales which should
be used in the model. There are four time scales appearing in the model:

\begin{itemize}
\item proper time of the observer (satellite): $\tau_o$.

\item proper time of the $i$th tracking station:
$\tau^{(i)}_{\rm station}$.

\item coordinate time $t=\TCB$ of the BCRS (alternatively a scaled
version of \TCB\ called \TDB\ can be used: $\TDB=(1-L_B)\,\TCB$ with
the current best estimate of the scaling constant $L_B\approx
1.55051976772\cdot10^{-8}\pm2\cdot10^{-17}$
\citep{Irwin:Fukushima:1999,IAU:2001}).

\item coordinate time $T=\TCG$ of the GCRS (alternatively a scaled
version of \TCG\ called \TT\ can be used: $\TT=(1-L_G)\,\TCG$,
$L_G\equiv6.969290134\cdot10^{-10}$ being a defining constant
\citep{IAU:2001}).

\end{itemize}

\noindent
It is clear that the observational data (e.g. in the case of the scanning
satellites like HIPPARCOS, GAIA and DIVA these are the projections of
the vector $\ve{s}$ on a local reference system of the satellite
which rotates together
with the satellite) are parametrized by the proper time of the satellite
$\tau_o$. It is also clear that the final catalog containing positions,
parallaxes and proper motions of the sources relative to the BCRS
should be parametrized by \TCB. The other two time scales (proper times
of the tracking station(s) $\tau^{(i)}_{\rm station}$ and \TCG) are used
exclusively for orbit determination.

The transformation between the proper time of the satellite $\tau_o$ and
\TCB\ can be done by integrating the equation

\begin{equation}\label{tau-s-TCB}
{d\,\tau_o\over dt}=
1-{1\over c^2}\left(
{1\over2}\,\dot{\ve{x}}_o^2+w(\ve{x}_o)
\right)+\OO4,
\end{equation}

\noindent
where $\ve{x}_o$ and $\dot{\ve{x}}_o$ are the BCRS position and
velocity of the satellite and $w(\ve{x}_o)$ is the gravitational
potential of the Solar system which can be approximated by

\begin{equation}\label{w(x-s)}
w(\ve{x}_o)\approx\sum_A {GM_A\over |\ve{r}_{oA}|},
\end{equation}

\noindent
where $\ve{r}_{oA}=\ve{x}_o-\ve{x}_A$, $M_A$ is the mass of body $A$,
and $\ve{x}_A=\ve{x}_A(t)$ is its barycentric position. Both
higher-order multipole moments of all the bodies and additional
relativistic terms are neglected in (\ref{w(x-s)}). The transformation
between the proper time of a tracking station and \TCG\ can be performed
in a similar way. The transformation between \TCG\ and \TCB\ is given in
the IAU Resolutions B1.3 (general post-Newtonian expression) and B1.5
(an expression for the accuracy of $5\cdot10^{-18}$ in rate and 0.2 ps
in amplitude of periodic effects) of \citet{IAU:2001}. There are
several analytical and numerical formulas for the position-independent
part of the transformation \citep[see, e.g.][and reference cited
therein]{Fukushima:1995,Irwin:Fukushima:1999}.

Although the use of the relativistic time scales described above is
indispensable from the theoretical and conceptual points of view, from
a purely practical point of view considerations of accuracy can be used
here to simplify the model. However, this depends on the particular
parameters of the mission and will be not analyzed here. In the
following it is assumed that the observed directions $\ve{s}$ are given
together with the corresponding epochs of observation $t_o$ in \TCB\
scale.

\section{Motion of the satellite}
\label{Section-motion}

It is well known that in order to compute the Newtonian aberration with
an accuracy of 1 \muas\ one needs to know the velocity of the observer
with an accuracy of $\sim 10^{-3}$ m/s \citep[see, e.g.][]{GAIA:2000}.
This is a rather stringent requirement and special care must be taken
to attain such an accuracy. Modeling of the satellite motion with such
an accuracy is a difficult task involving complicated equations of
motion which take into account various non-relativistic (Newtonian
$N$-body force, radiation pressure, active satellite thrusters, etc.)
as well as relativistic effects. Here some general recipe concerning
the relativistic part of the modeling will be given. Both the
non-relativistic parts of the model and a detailed study of the
relativistic effects in the satellite motion are beyond the scope of
the present paper.

In the relativistic model of positional observations developed in the
following Sections it is assumed that the observations are performed
from a space station or an Earth satellite whose position $\ve{x}_o$
relative to the BCRS is known for any moment of barycentric coordinate
time $t$. For those satellites the orbits of which are not located in
the vicinity of the Earth (this is the case for both GAIA and SIM) it
is advantageous to model their motion directly in the BCRS. Since the
mass of the satellite is too small to noticeably affect the motion of
other bodies of the Solar system, the equations of geodetic motion in
the BCRS can be used as equations of motion of the satellite. It is
sufficient (at least in the relativistic part of the equations) to
neglect here the multipole structure of the gravitating bodies as well
as the gravitational field produced by the rotational motion of these
bodies. Therefore, considering a system of $N$ bodies, each of which
can be characterized by position $\ve{x}_A$, velocity $\dot{\ve{x}}_A$
and mass $M_A$, the equations of motion of the satellite read
\citep{Will:1993,Klioner:Soffel:2000}

\begin{eqnarray}\label{mono-aEi}
{d^2\over dt^2}\, \ve{x}_o&=&-\sum_{A} G\,{\cal M}_A\,
{\ve{r}_{oA}\over |\ve{r}_{oA}|^3}
\nonumber\\
&&+{1\over c^2}\,\sum_{A} G\,{\cal M}_A\,
{\ve{r}_{oA}\over |\ve{r}_{oA}|^3}\,
\Biggl\{
(2\beta-1) \sum_{B\ne A} {G\,{\cal M}_B\over |\ve{r}_{AB}|}
\nonumber\\
&&\phantom{+{1\over c^2}\,\sum_{A} G\,{\cal M}_A\,
{\ve{r}_{oA}\over |\ve{r}_{oA}|^3}\,
\Biggl\{ }
+2(\gamma+\beta) \sum_{B} {G\,{\cal M}_B\over |\ve{r}_{oB}|}
\nonumber \\
&&
\phantom{
+{1\over c^2}\,\sum_{A} G\,{\cal M}_A\,
{\ve{r}_{oA}\over |\ve{r}_{oA}|^3}\,
\Biggl\{ }
+{3\over 2} {{\left(\ve{r}_{oA}\cdot \dot{\ve{x}}_A\right)}^2\over |\ve{r}_{oA}|^2}
\nonumber\\
&&
\phantom{
+{1\over c^2}\,\sum_{A} G\,{\cal M}_A\,
{\ve{r}_{oA}\over |\ve{r}_{oA}|^3}\,
\Biggl\{ }
-{1\over 2}\sum_{B\ne A} G\,{\cal M}_B\,
{\ve{r}_{oA}\,\cdot\,\ve{r}_{AB}\over |\ve{r}_{AB}|^3}
\nonumber\\
&&
\phantom{
+{1\over c^2}\,\sum_{A} G\,{\cal M}_A\,
{\ve{r}_{oA}\over |\ve{r}_{oA}|^3}\,
\Biggl\{ }
-(1+\gamma)\,\dot{\ve{x}}_A\,\cdot\,\dot{\ve{x}}_A
-\gamma\,\dot{\ve{x}}_o\,\cdot\,\dot{\ve{x}}_o
\nonumber \\
&&
\phantom{
+{1\over c^2}\,\sum_{A} G\,{\cal M}_A\,
{\ve{r}_{oA}\over |\ve{r}_{oA}|^3}\,
\Biggl\{ }
+2(1+\gamma)\,\dot{\ve{x}}_o\,\cdot\,\dot{\ve{x}}_A
\Biggr\}
\nonumber\\
&&+{1\over c^2}\,\sum_{A} G\,{\cal M}_A\,
{\dot{\ve{x}}_o-\dot{\ve{x}}_A
\over |\ve{r}_{oA}|^3}\,
\biggl\{
 2(1+\gamma)\,\dot{\ve{x}}_o\,\cdot\,\ve{r}_{oA}
-(2\gamma+1)\,\dot{\ve{x}}_A\,\cdot\,\ve{r}_{oA}\biggr\}\,
\nonumber\\
&&-{1\over c^2}\,\left(2\gamma+{3\over 2}\right)\,
\sum_{A}{G\,{\cal M}_A\over |\ve{r}_{oA}|}\,
\sum_{B\ne A} G\,{\cal M}_B\,{\ve{r}_{AB} \over |\ve{r}_{AB}|^3}+\OO4,
\end{eqnarray}

\noindent
where $\ve{r}_{oA}=\ve{x}_o-\ve{x}_A$, $\ve{r}_{AB}=\ve{x}_A-\ve{x}_B$,
and a dot denotes the total time derivative with respect to $t=\TCB$:
e.g. $\dot{\ve{x}}_o={d\over dt}\,\ve{x}_o(t)$.

The observations of the satellite itself performed to determine its
orbit (typically range and Doppler tracking) should be also modeled in
the relativistic framework: (1) the positions of the observing stations
should be defined in the GCRS and then transformed into the BCRS, (2)
the description of the signal propagation between the satellite and the
observing stations must take into account the corresponding
relativistic effects in the BCRS (the Shapiro effect, the relativistic
Doppler effects, etc.), (3) the difference between the proper time
scales of the satellite and the observing sites and the coordinate time
of the BCRS must also be taken into account. A detailed description of
these steps is beyond the scope of the present paper. The final result
of the orbit determination procedure is the BCRS position $\ve{x}_o$
and velocity $\dot{\ve{x}}_o$ of the satellite as a function of
$t=\TCB$.

If the satellite is close enough to the Earth (geostationary or lower)
it is advantageous to model its motion in the GCRS. The structure of
the equations of motion of a satellite with respect to the GCRS can be
written as

\begin{eqnarray}\label{ppN-eqm-struct}
{d^2\over dT^2}\,\ve{X}_o&=&\ve{\Phi}_{\rm E}+\ve{\Phi}_{\rm el}
+{1\over c^2}\left(
\ve{\Phi}_{\rm coup}+
\ve{\Phi}_{\rm mg}+
\ve{\Phi}_{\rm SEP}
\right)+\OO4,
\end{eqnarray}

\noindent
where $\ve{\Phi}_{\rm E}$ is the acceleration due to the gravitational
field of the Earth, $\ve{\Phi}_{\rm coup}$ is the Earth-third body
coupling term, $\ve{\Phi}_{\rm el}$ is the ``gravito-electric'' part
(independent of the velocity of the satellite) of the tidal influence
of external bodies (this formally includes the Newtonian tidal force),
$\ve{\Phi}_{\rm mg}$ is the purely relativistic ``gravito-magnetic''
part (depending on the velocity of the satellite) of the tidal
influence of external bodies, and, finally, $\ve{\Phi}_{\rm SEP}$ is
the additional acceleration due to possible violation of the strong
equivalence principle in alternative theories of gravity
($\ve{\Phi}_{\rm SEP}\equiv0$ in general relativity where the PPN
parameters $\beta=\gamma=1$). The main relativistic terms in
$\ve{\Phi}_{\rm E}$ come from the spherically symmetric part of the
Earth's gravitational field and can be formally derived in the
framework of the Schwarzschild solution for the Earth considered to be
isolated. This main relativistic effect is recommended to be taken into
account in the IERS Conventions \citep[][, Chapter 11]{IERS:1996}.
Explicit formulas
for the right-hand side of (\ref{ppN-eqm-struct}) can be found in
Section VIII of \citet{Klioner:Soffel:2000} in the framework of the PPN
formalism with parameters $\beta$ and $\gamma$. The same equations in
the framework of general relativity were derived by
\citet{Klioner:Voinov:1993}, \citet{Damour:Soffel:Xu:1994} and by
\citet{Brumberg:Kopeikin:1989b}.

If Eq. (\ref{ppN-eqm-struct}) is used to represent the motion of the
satellite, the whole process of the orbit determination can be
performed directly in the GCRS. Again the orbit determination
observations should be consistently modeled in the relativistic
framework: both the coordinates of the stations and the position of the
satellite should be described in the GCRS, the propagation of the
electromagnetic signals should be adequately modeled in the GCRS, the
proper time of the satellite, that of the tracking stations and \TCG\
should be properly converted into each other when needed. The result of
the orbit determination process is the GCRS coordinates of the
satellite $\ve{X}_o$ and its velocity ${d\over dT}\,\ve{X}_o$ as a
function of $T=\TCG$. In order to be used in the model of positional
observations given below these coordinates must be transformed into the
corresponding BCRS coordinates
\citep[see][]{IAU:2001,Klioner:Soffel:2000}:

\begin{equation}\label{Xa-xi}
\ve{X}_o=\ve{r}_{oE}+{1\over c^2}\left(
{1\over 2}\,\ve{v}_E\,(\ve{v}_E\cdot\ve{r}_{oE})+\gamma\,w_{\rm ext}(\ve{x}_E)\,
\ve{r}_{oE}+
\gamma\,\ve{r}_{oE}\,(\ve{a}_E\cdot\ve{r}_{oE})
-{1\over2}\,\gamma\,\ve{a}_E\,|\ve{r}_{oE}|^2
\right)+\OO4,
\end{equation}

\noindent
where $\ve{r}_{oE}=\ve{x}_o-\ve{x}_E$, and $\ve{x}_E$, $\ve{v}_E$ and
$\ve{a}_E$ are the BCRS position, velocity and acceleration of the
Earth, and $w_{\rm ext}(\ve{x}_E)$ is the gravitational potential of
the Solar system except for that of the Earth evaluated at the
geocenter. It is easy to estimate that the relativistic effects in
(\ref{Xa-xi}) may amount to $\sim 1$ m for $|\ve{r}_{oE}|\sim50000$ km. The
GCRS velocity of the satellite can be transformed into the
corresponding BCRS velocity as (this formula can be derived by taking
time derivative of (\ref{Xa-xi}) and using the \TCG-\TCB\
transformation \citep{IAU:2001,Klioner:Soffel:2000})

\begin{eqnarray}\label{Va-deltavi}
{d\over dT}\,\ve{X}_o=\delta \ve{v}_o+{1\over c^2}\biggl(
&&\delta \ve{v}_o\,\left({1\over 2}\,|\ve{v}_E|^2
+(1+\gamma)\,(w_{\rm ext}(\ve{x}_E)+\ve{a}_E\cdot\ve{r}_{oE})
+\ve{v}_E\cdot\delta\ve{v}_o\right)
\nonumber\\
&&
+{1\over2}\,\ve{v}_E\,(\ve{v}_E\cdot\delta\ve{v}_o)
+\gamma\,\ve{r}_{oE}\,(\ve{a}_E\cdot\delta\ve{v}_o)
-\gamma\,\ve{a}_E\,(\ve{r}_{oE}\cdot\delta\ve{v}_o)
\nonumber\\
&&
+{1\over2}\,\ve{a}_E\,(\ve{v}_E\cdot\ve{r}_{oE})
+{1\over2}\,\ve{v}_E\,(\ve{a}_E\cdot\ve{r}_{oE})
+\gamma\,{\dot w}_{\rm ext}(\ve{x}_E)\,\ve{r}_{oE}
\nonumber\\
&&
+\gamma\,\ve{r}_{oE}\,(\dot{\ve{a}}_E\cdot\ve{r}_{oE})
-{1\over2}\,\gamma\,\dot{\ve{a}}_E\,|\ve{r}_{oE}|^2
\biggr)
+\OO4,
\end{eqnarray}

\noindent
where $\delta \ve{v}_o=\dot{\ve{x}}_o-\ve{v}_E$ and
$\dot{\ve{a}}_E={d\over dt}\ve{a}_E$. It is easy to see from
(\ref{Va-deltavi}) that for a geostationary satellite numerical
difference between ${d\over dT}\,\ve{X}_o$ and $\delta \ve{v}_o$ is
less than $10^{-4}$ m/s. Therefore, for the model with a final accuracy
of $\sim 1$ \muas\ the relativistic terms in (\ref{Va-deltavi}) can be
neglected, and one can use ${d\over dT}\,\ve{X}_o=\delta \ve{v}_o$.

Note that probably even for the goal accuracy of $10^{-3}$ m/s in the velocity
of the satellite one can significantly simplify both the BCRS equations
of motion (\ref{mono-aEi}) and those in the GCRS
(\ref{ppN-eqm-struct}). However, this crucially depends on the
particular orbit of the satellite and such an analysis will not be
given here.

Let us also note that the rotational motion of the satellite should
also be carefully modeled (for some missions the attitude of the
satellite will be determined from the observational data produced by
the satellite, but nevertheless a kind of theoretical modeling is still
necessary). From the theoretical point of view in order to model the
rotational motion of the satellite it is convenient to introduce a
local kinematically-nonrotating reference system for the satellite with
coordinates $(\tau_o,\ve{\xi})$, where $\tau_o$ is the proper time of
the satellite related to $t$ by (\ref{tau-s-TCB}) and $\ve{\xi}$ are
the spatial coordinates of the reference system related to those of
BCRS by a transformation analogous to (\ref{Xa-xi}) \citep[see][for
details on the local reference system of the satellite]
{Brumberg:Kopeikin:1989a,Klioner:1993,Klioner:Soffel:1998a}. The
rotational motion of the satellite with respect to the axes $\ve{\xi}$
is described by the post-Newtonian rotational equations of motion
discussed by \citet{Damour:Soffel:Xu:1993} for the case of general
relativity and by \citet{Klioner:Soffel:1998b,Klioner:Soffel:2000} in
the framework of the PPN formalism. From the practical point of view,
it is, however, clear that the largest relativistic effect in the
rotational motion of the satellite with respect to remote stars is due
to the geodetic (de Sitter) precession which amounts to $\sim 2$ \muas\
per hour ($\sim2\arcsec$ per century) for the satellites on a
heliocentric orbit with the semi-major axis close to 1 AU (like GAIA
and SIM) and $\sim140$ \muas\ per hour ($\sim 1.2\arcsec$ per year) for
the satellites on a geostationary orbit (like HIPPARCOS or DIVA). Taking
into account that the satellites will typically monitor and verify
their attitude with the help of onboard gyroscopes {\it and}
observations of specially selected stars, and that the precise attitude
with respect to remote stars will be determined aposteriori from the
processing of observational data, it is unlikely that such small
relativistic effects could be of practical importance.

The rest of the relativistic model of positional observations is
totally independent of the GCRS. It is only the orbit determination
process (or at least a part of that process) which involves the use of
the GCRS coordinates and concepts.

\section{Aberration}
\label{Section-aberration}

The first step of the model is to get rid of the aberrational effects
induced by the barycentric velocity of the observer. Let $\ve{s}$
denote the unit direction ($\ve{s}\,\cdot\,\ve{s}=1$) {\sl toward} the
source as observed by the observer (satellite). Let $\ve{p}$ be the
BCRS coordinate velocity of the photon at the point of observation.
Note that $\ve{p}$ is directed roughly from the source to the observer. The
unit BCRS coordinate velocity of the light ray $\ve{n}=\ve{p}/|\ve{p}|$
(this implies $\ve{n}\,\cdot\,\ve{n}=1$) can be then computed as
\citep{Klioner:1991b,Klioner:Kopeikin:1992}

\begin{eqnarray}\label{aberration}
\ve{s}=-\ve{n}&+&\,{1\over c}\,\ve{n}\,\times\,(\dot{\ve{x}}_o\,\times\,\ve{n})
\nonumber\\
&+&{1\over c^2}\,\Biggl\{\,(\ve{n}\,\cdot\,\dot{\ve{x}}_o)\
\ve{n}\,\times\,(\dot{\ve{x}}_o\,\times\,\ve{n})
+{1\over 2}\,\dot{\ve{x}}_o\,\times\,(\ve{n}\,\times\,\dot{\ve{x}}_o)
\,\Biggr\}
\nonumber\\
&+&{1\over c^3}\,\Biggl\{
\left({(\ve{n}\,\cdot\,\dot{\ve{x}}_o)}^2+(1+\gamma)\,w(\ve{x}_o)\right)\
\ve{n}\,\times\,(\dot{\ve{x}}_o\,\times\,\ve{n})
+{1\over 2}\,(\ve{n}\,\cdot\,\dot{\ve{x}}_o)\
\dot{\ve{x}}_o\,\times\,(\ve{n}\,\times\,\dot{\ve{x}}_o)
\Biggr\}
\nonumber\\
&+&\OO4,
\end{eqnarray}

\noindent
where $w(\ve{x}_o)$ is the gravitational potential of the Solar system
at the point of observation.  This formula contains relativistic
aberrational effects up to the third order with respect to $1/c$. One
can check from (\ref{aberration}) that condition
$\ve{n}\,\cdot\,\ve{n}=1$ implies $\ve{s}\,\cdot\,\ve{s}=\OO4$. Because
of the first-order aberrational terms (classical aberration) in order
to attain the accuracy of 1 $\mu$as the BCRS coordinate velocity of the
satellite must be known to an accuracy of $\sim10^{-3}$ m/s. For a
satellite with the BCRS velocity $|\dot{\ve{x}}_o|\sim40$ km/s, the
first-order aberration is of the order of 28\arcsec, the second-order
one may amount to 3.6 mas, and the third-order effects are $\sim 1$
\muas. Note also that the higher-order aberrational effects are
nonlinear with respect to the velocity of the satellite and cannot be
divided into additive pieces like ``annual" and ``diurnal'' aberrations
as it could be done with the first-order aberration for an Earth-bound
observer. This is the reason why it is advantageous to use directly the
BCRS velocity of the satellite.

Vector $\ve{s}$ is defined in the kinematically non-rotating local
satellite reference system $(\tau_o,\ve{\xi})$ which is mentioned in
the previous Section. As it was discussed in Section 7.1 of
\citet{Klioner:Kopeikin:1992} this point of view is equivalent to the
standard tetrad approach discussed, e.g. by \citet{Brumberg:1986}. The
same form of Eq. (\ref{aberration}) can be derived both from the
standard tetrad formalism \citep[see,
e.g.][]{Brumberg:1986,Klioner:1991b} and from the considerations
related to the local reference system of the satellite
\citep{Klioner:Kopeikin:1992}. Actual observations of a scanning
astrometric satellite are referred to a reference system, spatial axes
of which rigidly rotate with respect to $\ve{\xi}$ so that the
satellite's attitude remains fixed in the rotating axes. The rotating
axes $\overline{\ve{\xi}}$ are related to the kinematically
non-rotating axes $\ve{\xi}$ as

\begin{equation}\label{rotating-vs-nonrotating-SRS}
\overline{\xi}^a=P^{ab}(\tau_o)\,\xi^b,
\end{equation}

\noindent
where $P^{ab}$ is a time-dependent orthogonal matrix, which can be
parametrized, e.g. by three Euler angles. These Euler angles define
the attitude of the satellite with respect to the kinematically
non-rotating axis and should be determined from observations and
the corresponding modeling. Depending on the particular optical scheme
of the scanning satellite rotational motion of the satellite can
produce an additional measurable aberrational effect. A detailed
analysis of this additional effect will be published
elsewhere.

To clarify the origin of the terms in (\ref{aberration}) proportional
to the gravitational potential $w$ let us consider a fictitious
observer, whose position coincides with that of the satellite at the
moment of observation and whose velocity with respect to the BCRS is
zero. The direction toward the source observed by this fictitious
observer

\begin{equation}\label{s-prim}
\ve{s}^\prime=-\ve{n}+\OO4
\end{equation}

\noindent
as can be calculated from
(\ref{aberration}) for $\dot{\ve{x}}_o=0$.  On the other hand, the
transformation between the directions toward the source observed by the
two observers located at the same point of space-time can be derived
from the usual Lorentz transformation which can be used in its closed
form to speed up the practical calculations (F.~Mignard 2000, private
communication). The Lorentz transformation depends only on the velocity
of one observer as seen by the other one $\ve{v}$. Hence, using the
Lorentz transformation in its closed form one gets

\begin{eqnarray}\label{aberration-Lorentz}
\ve{s}&=&\left(\ve{s}^\prime+
\left\{\,{\Gamma\over c\,}+
\left[\,\Gamma-1\,\right]\,
{\ve{v}\cdot\ve{s}^\prime\over |\ve{v}|^2}\,\right\}\,\ve{v}\,
\right)\,
{1\over \Gamma\,\left(1+\ve{v}\cdot\ve{s}^\prime/c\right)},
\\ \label{Gamma}
\Gamma&=&{1\over \sqrt{1-|\ve{v}|^2/c^2}}.
\end{eqnarray}

\noindent
It is easy to show from (\ref{aberration-Lorentz}) that
$\ve{s}^\prime\,\cdot\,\ve{s}^\prime=1$ implies $\ve{s}\,\cdot\,\ve{s}=1$. The
velocity of the satellite as measured by the fictitious observer
$\ve{v}$ can again be calculated either with the help of the tetrad
formalism \citep[e.g.][]{Brumberg:1986} or by considering the local
reference system for the fictitious observer
\citep{Klioner:Kopeikin:1992,Klioner:1993}. The latter line of
arguments leads to an equation similar to (\ref{Va-deltavi}) where one
should put to zero the BCRS velocity of the fictitious observer (this
corresponds to $\ve{v}_E=0$ in (\ref{Va-deltavi})), its BCRS
acceleration ($\ve{a}_E=0$ in (\ref{Va-deltavi})) as well as the
position of the satellite relative to the local reference system
($\ve{r}_{oE}=0$ in (\ref{Va-deltavi})). Therefore, one gets

\begin{equation}\label{velocity-renormalization}
\ve{v} = \dot{\ve{x}}_o\,
\left(\,1+{1\over c^2}\,(1+\gamma)\,w(\ve{x}_o)\,\right)+\OO4.
\end{equation}

\noindent
It is this renormalization of the velocity of the satellite which leads
to the $w$-dependent terms in (\ref{aberration}). It is easy to check
that formulas
(\ref{s-prim})--(\ref{velocity-renormalization}) are
equivalent to (\ref{aberration}).

Alternatively, it is easy to calculate a formula relating the observed
angular distance $\varphi_{12}^{\rm obs}$ ($\cos\varphi_{12}^{\rm
obs}=\ve{s}_1\cdot\ve{s}_2$) between two given sources to the
coordinate angular distance $\varphi_{12}^{\rm coord}$ between them
($\cos\varphi_{12}^{\rm coord}=\ve{n}_1\cdot\ve{n}_2$):

\begin{eqnarray}\label{aberration-cos}
\cos\varphi_{12}^{\rm obs}=
\cos\varphi_{12}^{\rm coord}
+\left(\cos\varphi_{12}^{\rm coord}-1\right)&
\Biggl[&
-\left(1+{1\over c^2}\,(1+\gamma)\,w(\ve{x}_o)\right)\,{|\dot{\ve{x}}_o|\over c}\,
(\cos\theta_1+\cos\theta_2)
\nonumber\\
&&
+{|\dot{\ve{x}}_o|^2\over c^2}\,
(\cos^2\theta_1+\cos^2\theta_2+\cos\theta_1\,\cos\theta_2-1)
\nonumber\\
&&
-{|\dot{\ve{x}}_o|^3\over c^3}\,
(\cos\theta_1+\cos\theta_2)\,
(\cos^2\theta_1+\cos^2\theta_2-1)
\Biggr]
\nonumber\\
&&
+\OO4,
\end{eqnarray}

\noindent
where $\theta_i$ are the angles between the direction toward the $i$th
source and the direction of the velocity $\dot{\ve{x}}_o$:

\begin{eqnarray}\label{psi-i}
\cos\theta_i&=&-{1\over |\dot{\ve{x}}_o|}\ve{n}_i\,\cdot\,\dot{\ve{x}}_o,\qquad i=1,2.
\end{eqnarray}

The terms in (\ref{aberration}) and
(\ref{aberration-cos}) proportional to $w$ are the largest terms of the
third order. The third-order terms are of order 1 \muas\
and, therefore, the potential $w$ can be approximated here by the
potential of the spherically symmetric Sun

\begin{eqnarray}\label{U-sun}
w(\ve{x}_o)\approx{GM_{\rm sun}\over |\ve{x}_o-\ve{x}_{\rm Sun}|}.
\end{eqnarray}


\noindent
A closed-form expression equivalent to (\ref{aberration-cos}) can be
derived from (\ref{aberration-Lorentz}):

\begin{eqnarray}\label{aberration-cos-Lorenz}
1-\cos\varphi_{12}^{\rm obs}&=&
\left(1-\cos\varphi_{12}^{\rm coord}\right)\,
{1-\displaystyle{\left|\ve{v}\right|^2\over c^2} \over
\displaystyle\left(1+{\left|\ve{v}\right|\over c}\,\cos\theta_1\right)
\displaystyle\left(1+{\left|\ve{v}\right|\over c}\,\cos\theta_2\right)},
\end{eqnarray}

\noindent
where $\ve{v}$ is defined by (\ref{velocity-renormalization}).

For order-of-magnitude estimations it is also useful to derive a
formula for the angular shift $\delta\theta$ of the source toward the
apex of the satellite's motion:

\begin{eqnarray}\label{aberation-angle}
\delta\theta&=&{1\over c}\,|\dot{\ve{x}}_o|\,\sin\theta\
\left(1+{1\over c^2}\,(1+\gamma)\,w(\ve{x}_o)+{1\over 4}\,{|\dot{\ve{x}}_o|^2\over c^2}
\right)
\nonumber\\
&&
-{1\over 4}\,{|\dot{\ve{x}}_o|^2\over c^2}\,\sin2\theta
+{1\over 12}\,{|\dot{\ve{x}}_o|^3\over c^3}\,\sin3\theta
+\OO4
\end{eqnarray}

\noindent
or in the closed form

\begin{eqnarray}\label{aberation-angle-Lorenz}
\cos(\theta-\delta\theta)=
{{\cos\theta+\displaystyle{\left|\ve{v}\right|\over c}}
\over
{1+\displaystyle{\left|\ve{v}\right|\over c}\,\cos\theta}}.
\end{eqnarray}

\noindent
Here $\theta$ is the angle between the direction toward the source and
the direction of the satellite's velocity
$\cos\theta=-\ve{n}\,\cdot\,\ve{v}\,/\,|\ve{v}|=-\ve{n}\,\cdot\,\dot{\ve{x}}_o\,/\,|\dot{\ve{x}}_o|$.

\section{Gravitational light deflection}
\label{Section-gravitational-deflection}

The next step is to account for the gravitational
light defection, that is to convert $\ve{n}$  into the corresponding
unit coordinate direction $\ve{k}$ from the observed source at the
moment of emission to the observer at the moment of observation. Here
two different cases should be distinguished: (1) objects outside of the
Solar system for which their finite distance from the origin of the
BCRS plays [almost] no role for this step of the reduction scheme, and
(2) Solar system objects, the finite distance of which must be taken
into account. Let us first relate $\ve{n}$ to the unit coordinate
direction $\vecg{\sigma}$ of the light propagation infinitely far from
the gravitating sources for $t\to-\infty$ and then consider the
influence of the finite distance to the objects on the gravitational
light deflection separately for these two classes of objects (Sections
\ref{Subsection-coupling-solar-system} and
\ref{Subsection-coupling-stars} below).

The Solar system is assumed here to be isolated. This means that the
gravitational field produced by the matter outside of the Solar system
is neglected.  This assumption is well founded if the time dependence of
the gravitational field produced outside of the Solar system is
negligible. Otherwise the external gravitational field must be
explicitly taken into account (e.g. for observations of edge-on binary
stars, where the gravitational field of the companion can cause an
additional time-dependent light deflection, or for astrometric
microlensing events). Some of such cases will be discussed below in
Section \ref{Section-beyond-the-model}.

The equations of light propagation can be derived from the
general-relativistic Maxwell equations \citep{Misner:Thorne:Wheeler:1973}.
It is sufficient, however, to
consider only the limit of geometric optics. The relativistic effects
depending on wavelength (and therefore representing deviations from
geometric optics) are much smaller than 1 \muas\ in the Solar system
\citep[see, e.g.][]{Mashhoon:1974b}. In the limit of geometric optics
the relativistic equations of light propagation can be written in the
form

\begin{eqnarray}\label{x-p}
\ve{x}_p(t)=\ve{x}_p(t_o)+c\,\vecg{\sigma}\,(t-t_o)+\Delta\ve{x}_p(t),
\end{eqnarray}

\noindent
where $t_o$ is the moment of observation, $\ve{x}_p(t_o)$ is the
position of the photon at the moment of observation (this position
obviously coincides with the position of the satellite at that moment
$\ve{x}_p(t_o)=\ve{x}_o(t_o)$), $\vecg{\sigma}$ is the unit coordinate
direction of the light propagation at the past null infinity
($\vecg{\sigma}\cdot\vecg{\sigma}=1$)

\begin{eqnarray}\label{sigma}
\vecg{\sigma}=\lim_{t\to-\infty} {1\over c}\,\dot{\ve{x}}_p(t),
\end{eqnarray}

\noindent
and $\Delta\ve{x}_p$ is the sum of all gravitational effects in the
light propagation which satisfies the conditions

\begin{eqnarray}\label{condition1}
\Delta\ve{x}_p(t_o)&=&0,
\\
\label{condition2}
\lim\limits_{t\to-\infty}\Delta\dot{\ve{x}}_p(t)&=&0.
\end{eqnarray}

\noindent
From the BCRS metric \citep{IAU:2001} one can easily see that
$\Delta\ve{x}_p(t)\sim{\cal O}(c^{-2})$ and ${1\over
c}\,\Delta\dot{\ve{x}}_p(t)\sim{\cal O}(c^{-2})$.

Several effects in $\Delta\ve{x}_p$ should be apriori considered at the
level of 1 \muas: (1) the effects of the spherically symmetric part of
the gravitational field of each sufficiently large gravitating body,
(2) the effects due to the non-sphericity (mainly due to the quadrupole
moments) of the bodies, (3) the effects caused by the gravitomagnetic
field due to the translational motion of the bodies, (4) the effects
caused by the gravitomagnetic field due to the rotational motion of the
bodies. The reduction formulas for all there effects have been derived
and discussed in detail by
\citet{Brumberg:Klioner:Kopejkin:1990,Klioner:1991a,Klioner:1991b,Klioner:Kopeikin:1992}.

Table \ref{Table-gravitational-deflection} illustrates the maximal
magnitudes of the various gravitational effects due to the Solar system
bodies and the maximal angular distances between the source and the
body at which the gravitational light deflection from that body should
still be accounted for to attain the final accuracy of 1 \muas. Note
that the values in Table \ref{Table-gravitational-deflection} are
slightly different from those published by
\citet{Brumberg:Klioner:Kopejkin:1990,Klioner:2000}. The reason for
this discrepancy is that for Table \ref{Table-gravitational-deflection}
the best current estimates for the physical parameters of the Solar
system bodies were taken from \citet{Weissman:et:al:1999}, whereas the
IAU (1976) system of astronomical constants was used for the previous
publications.


One can see that the post-post-Newtonian terms attain 1 \muas\ only
within 53\arcmin\ from the Sun and can be currently neglected in the
case of space astrometry since none of the proposed satellites could
observe so close to the Sun. For the same reason the effect due to the
rotational motion of the Sun amounting to 0.7 \muas\ for a grazing ray
is unobservable. The largest observable effect due to the rotational
motion is 0.2 \muas\ for a light ray grazing Jupiter. Therefore, the
effects due to the rotational motion of the bodies are also too small
to be taken into account.

The largest contribution in $\Delta\ve{x}_p$ for Solar system
applications comes from the spherically symmetric components of the
gravitational fields of the massive bodies and can be calculated as

\begin{eqnarray}\label{delta-x-p-schwarzschild}
\Delta_{pN}\ve{x}_p(t)&=&-\sum_A\,{(1+\gamma)GM_A\over c^2}\,
\biggl(\,\ve{d}_A\,{\cal I}_A\, +\vecg{\sigma}\, {\cal J}_A\biggr),
\\
\label{vector-d-A}
\ve{d}_A&=&\vecg{\sigma}\,\times\,(\ve{r}_{oA}\,\times\,\vecg{\sigma}),
\\
\label{I}
{\cal I}_A&=&{1\over |\ve{r}_A|-\vecg{\sigma}\cdot\ve{r}_A}-
  {1\over |\ve{r}_{oA}|-\vecg{\sigma}\cdot\ve{r}_{oA}},
\\
\label{J}
{\cal J}_A&=&\log { |\ve{r}_A|+\vecg{\sigma}\cdot\ve{r}_A\over
|\ve{r}_{oA}|+\vecg{\sigma}\cdot\ve{r}_{oA}},
\\
\label{R-A}
\ve{r}_A&=&\ve{x}_p(t)-\ve{x}_A,
\\
\label{R-A0}
\ve{r}_{oA}&=&\ve{x}_p(t_o)-\ve{x}_A=\ve{x}_o(t_o)-\ve{x}_A,
\end{eqnarray}

\noindent
so that

\begin{eqnarray}\label{delta-dot-x-p-schwarzschild}
{1\over c}\,\Delta_{pN}\dot{\ve{x}}_p(t)&=&-\sum_A\,{(1+\gamma)GM_A\over c^2}\,
\biggl(\,\ve{d}_A\,{1\over c}\,
\dot{\cal I}_A\,+\vecg{\sigma}\,{1\over c}\,\dot{\cal J}_A\biggr),
\\
\label{I-dot}
{1\over c}\,\dot{\cal I}_A&=&{1\over |\ve{r}_A|\,(|\ve{r}_A|-\vecg{\sigma}\cdot\ve{r}_A)},
\\
\label{J-dot}
{1\over c}\,\dot{\cal J}_A&=&{1\over |\ve{r}_A|},
\end{eqnarray}

\noindent
where $\ve{x}_A$ is the position and $M_A$ is the mass of body A.

The positions of the bodies $\ve{x}_A$ are supposed to be constant in
(\ref{delta-x-p-schwarzschild})--(\ref{J-dot}). In reality, however,
the bodies are moving and this motion cannot be neglected in the
calculation of light deflection. It is hardly possible to calculate
analytically the light path in the gravitational field of an
arbitrarily moving body without resorting to some approximations
\citep[see, however,][]{Kopeikin:Schaefer:1999}. On the other hand,
numerical integration of the equations of light propagation which could
be used here as a remedy is too impractical for massive calculations
necessary in astrometry. Let us, therefore, consider a Taylor expansion
of $\ve{x}_A$:

\begin{equation}\label{x-A-expansion}
\ve{x}_A(t)=\ve{x}^{\rm eph}_A(t_{A0}) +\dot{\ve{x}}^{\rm eph}_A(t_{A0})\,(t-t_{A0})
+{\cal O}(\ddot{\ve{x}}^{\rm eph}_A).
\end{equation}

\noindent
Note that here $\ve{x}^{\rm eph}_A(t_{A0})$ and $\dot{\ve{x}}^{\rm
eph}_A(t_{A0})$ are the actual position and velocity of body $A$ taken
from an ephemeris for some moment $t_{A0}$, while $\ve{x}_A(t)$ is the model
for the position of body $A$ to be used for the approximate calculation
of the light propagation. One can integrate the equations of motion for
a light ray using (\ref{x-A-expansion}) for the coordinates of the
gravitating body. The solution was first derived by
\citet{Klioner:1989} and describes the light propagation in the
post-Newtonian approximation under the assumption that the gravitating
bodies move with a constant velocity (only the effects linear with
respect to velocity $\dot{\ve{x}}_A$ were taken into account, since
formally the terms quadratic in $\dot{\ve{x}}_A$ are of
post-post-Newtonian order $\OO4$). However, expansion
(\ref{x-A-expansion}) has a free parameter $t_{A0}$ which can be used
to minimize the error in the light propagation equations caused by the
higher-order terms neglected in (\ref{x-A-expansion}).
From an analysis of the residual terms in
the equations of light propagation proportional to the accelerations of
the bodies
\citet{Klioner:Kopeikin:1992}
have shown that one reasonable choice which guarantee the
residual effects to be negligible within the chosen approximation
scheme is to set $t_{A0}$ equal to the moment of the closest approach
$t^{\rm ca}_{A0}$ of the body and the photon

\begin{eqnarray}\label{moment-of-closest-approach}
t^{\rm ca}_{A0}&=&\max\left(t_e,t_o-
\max\left(0,{\ve{g}\cdot (\ve{x}_o(t_o)-\ve{x}^{\rm eph}_A(t_o))
             \over c\,|\ve{g}|^2}\right)
\right),
\\
\label{vector-g}
\ve{g}&=&\vecg{\sigma}-{1\over c}\,\dot{\ve{x}}^{\rm eph}_{A}(t_o),
\end{eqnarray}

\noindent
where $t_e$ is the time of emission of the light ray by the source (for
sources located outside of the Solar system one can put $t_e=-\infty$
so that the outer '$\max$' in (\ref{moment-of-closest-approach}) can be
omitted).

Recently an advanced formalism to integrate the equations of light
propagation in the field of arbitrarily moving gravitating bodies has
been developed by \citet{Kopeikin:Schaefer:1999}
and \citet{Kopeikin:Mashhoon:2002}. The authors suggest
to use the solution of the Einstein field equations in the form of retarded
potentials so that the positions of the gravitating bodies are computed
for the retarded moment of time

\begin{equation}\label{retarded-moment}
t^{\rm r}_{A0}=t_o-{1\over c}\,|\ve{x}_o(t_o)-\ve{x}^{\rm eph}_A(t^r_{A0})|,
\end{equation}

\noindent
and derive rigorous laws of the light propagation in the gravitational
field of arbitrarily moving point masses in the first post-Minkowskian
approximation (i.e. all the terms of order ${\cal O}(G^2)$ are
neglected). Using their rigorous approach the authors prove, however,
that if the positions and velocities of the bodies are calculated at
$t^{\rm r}_{A0}$ the effects due to the accelerations of the
gravitating bodies as well as those proportional to the second and
higher orders of velocities $\dot{\ve{x}}_{A}$ are much smaller than 1
\muas\ for Solar system applications and thereby completely negligible
for our model. On the other hand, if the effects due to the
accelerations and those of the second and higher orders with respect to
the velocities are neglected, so that each body is supposed to move with
a constant velocity the formulas for $\Delta\ve{x}_p(t)$
derived by \citep{Klioner:1989,Klioner:Kopeikin:1992} and those derived
by \citep{Kopeikin:Schaefer:1999} coincide. If the bodies are supposed
to move uniformly and rectilinearly and only the first-order terms with
respect of the velocities are taken into account, the resulting
solution for the light propagation can be considered as a numerical
approximation to the rigorous solution derived by
\citet{Kopeikin:Schaefer:1999}. The parameter $t_{A0}$ remains a free
parameter of this approximate solution. The approach due to
\citet{Kopeikin:Schaefer:1999} does not prove that the choice
$t_{A0}=t^{\rm r}_{A0}$ gives minimal residual terms in such an
approximate solution. It is not clear which moment $t^{\rm r}_{A0}$ or
$t^{\rm ca}_{A0}$ is numerically more advantageous to use in
(\ref{x-A-expansion}). However, if the effects explicitly proportional
to $\dot{\ve{x}}_A$ are taken into account to compute the light path,
the difference between the solutions for $t_{A0}=t^{\rm ca}_{A0}$ and
for $t_{A0}=t^{\rm r}_{A0}$ in the gravitational field of the Solar
system is several orders of magnitude lower than 1~\muas.

The effect explicitly proportional to $\dot{\ve{x}}_A$ has been
investigated in detail by
\citet{Klioner:1989}, \citet{Klioner:Kopeikin:1992} and
\citet{Kopeikin:Schaefer:1999}. If
the position of the body is taken for $t_{A0}=t^{\rm ca}_{A0}$ or
$t_{A0}=t^{\rm r}_{A0}$ this effect in the light deflection can be
estimated as $\delta_{T^\star}\sim {1\over
c}\,|\dot{\ve{x}}_A|\,\delta_{pN}$, where $\delta_{pN}$ is the
deflection induced by the spherically symmetric field of the body.
According to Table~\ref{Table-gravitational-deflection},
$\delta_{T^\star}$ may amount to 0.8 \muas\ for a light ray grazing Jupiter
and only 0.1 \muas\ in the case of the Sun. However, $\delta_{pN}$
attains its maximal value when the impact parameter of the light ray is
much smaller than the distances between the gravitating body and the
points of emission and observation. For this case one can prove that
$\delta_{T^\star}\sim {1\over c}\,
\left(\ve{\sigma}\cdot\dot{\ve{x}}_A\right)\,\delta_{pN}$. For the
special case of an observer situated near the Earth orbit the cosine
factor in $\ve{\sigma}\cdot\dot{\ve{x}}_A$ reduces the effects produced
by Jupiter and Saturn in their orbital motion by a factor of at least 4
and 6, respectively. Therefore, the effects explicitly proportional to
the velocities of the bodies can be neglected in our model.

If the effects due to the velocity of the body are completely
neglected, we effectively use a constant value

\begin{equation}\label{x-A-expansion-0}
\ve{x}_A(t)=\ve{x}_A(t_{A0})
\end{equation}

\noindent
for the coordinates of the bodies. Numerical simulations show that
in the gravitational field of the Solar system the
difference of the light deflection angle calculated using
$t_{A0}=t^{\rm ca}_{A0}$ and that using $t_{A0}=t^{\rm r}_{A0}$ in
(\ref{x-A-expansion-0}) does not exceed 0.01~\muas. Therefore, for
practical purposes both $t^{\rm ca}_{A0}$ and $t^{\rm r}_{A0}$ can be
used and $\ve{x}_A$ should be taken to be $\ve{x}_A(t^{\rm ca}_{A0})$
or $\ve{x}_A(t^{\rm r}_{A0})$ in (\ref{R-A})--(\ref{R-A0}) and all
related formulas. Note, however, that, e.g. the use of the moment of
reception $t_o$ instead of $t^{\rm ca}_{A0}$ or $t^{\rm r}_{A0}$ in
(\ref{x-A-expansion-0}) may lead to a significant error in the
calculated gravitational light deflection. This maximal error of this kind
can be roughly estimated as
$|\ve{r}_{oA}|/L\,\cdot\,|\dot{\ve{x}}_A|/c\,\cdot\,\delta_{pN}$, where
$L$ is the radius of the body and may amount to $\sim10$~mas in the
case of Jupiter.


It is to note that a number of smaller bodies should be also taken into
account. For a spherical body with the mean density $\rho$, the light
deflection is larger than $\delta$ if its radius

\begin{equation}\label{L-rho-delta}
L\ge \left(\rho\over 1\,{\rm g/cm^3}\right)^{-1/2}\,
     \left(\delta\over 1\,\muas\right)^{1/2}\,
     624\,{\rm km}.
\end{equation}

\noindent
Therefore, at the level of 1 \muas\ (and even 10 \muas) one should
additionally account for several largest satellites of the giant planets,
Pluto and
Charon, and also Ceres. The maximal values of the effects produced by
these bodies are evaluated in Table
\ref{Table-gravitational-deflection}. It is clear, however, that the
gravitational light deflection due to these small bodies could be
larger than 1 \muas\ only if a source is observed very close to the
corresponding body. The maximal angular distances at which this additional
gravitational deflection should be taken into account are also given in
Table \ref{Table-gravitational-deflection}.

Finally, the effects of the quadrupole fields of the giant planets should be
taken into account if the angular distance between the planet and the
object is smaller than the values given in the sixth column of Table
\ref{Table-gravitational-deflection}. The corresponding reduction
formulas for $\Delta_{Q}\ve{x}_p(t)$ and $\Delta_{Q}\dot{\ve{x}}_p(t)$
derived by \citet{Klioner:1991a} and \citet{Klioner:Kopeikin:1992} read

\begin{eqnarray}\label{delta-x-p-Q}
\Delta_{Q}\ve{x}_p(t)&=&
{1\over 2\,c^2}\,(1+\gamma)\,G\,\sum_A\,
\left(
\vecg{\alpha}_A\,{\cal U}_A
+\vecg{\beta}_A\,{\cal E}_A
+\vecg{\gamma}_A\,{\cal F}_A
+\vecg{\delta}_A\,{\cal V}_A
\right),
\\
\label{U}
{\cal U}_A&=&
{1\over |\ve{d}_A|}\,\left(
{1\over |\ve{r}_A|}\,
{|\ve{r}_A|+\vecg{\sigma}\cdot\ve{r}_A\over
 |\ve{r}_A|-\vecg{\sigma}\cdot\ve{r}_A}
-{1\over |\ve{r}_{oA}|}\,
{|\ve{r}_{oA}|+\vecg{\sigma}\cdot\ve{r}_{oA}\over
 |\ve{r}_{oA}|-\vecg{\sigma}\cdot\ve{r}_{oA}}
\right),
\\
\label{E}
{\cal E}_A&=&{\vecg{\sigma}\cdot\ve{r}_A\over |\ve{r}_A|^3}
-{\vecg{\sigma}\cdot\ve{r}_{oA}\over |\ve{r}_{oA}|^3},
\\
\label{F}
{\cal F}_A&=&|\ve{d}_A|\,\left({1\over |\ve{r}_A|^3}
-{1\over |\ve{r}_{oA}|^3}\right),
\\
\label{V}
{\cal V}_A&=&
-{1\over |\ve{d}_A|^2}\,\left(
{\vecg{\sigma}\cdot\ve{r}_A\over |\ve{r}_A|}
-{\vecg{\sigma}\cdot\ve{r}_{oA}\over |\ve{r}_{oA}|}\right),
\\
\label{ve-alpha}
\vecg{\alpha}_A&=&2\,\ve{f}_A-2\,(\ve{f}_A\cdot\vecg{\sigma})\,\vecg{\sigma}
-(\ve{g}_A\cdot\vecg{\sigma}+4\,\ve{f}_A\cdot\ve{h}_A)\,\ve{h}_A,
\\
\label{ve-beta}
\vecg{\beta}_A&=&2\,(\ve{f}_A\cdot\vecg{\sigma})\,\ve{h}_A
+(\ve{g}_A\cdot\vecg{\sigma}-\ve{f}_A\cdot\ve{h}_A)\,\vecg{\sigma},
\\
\label{ve-gamma}
\vecg{\gamma}_A&=&2\,(\ve{f}_A\cdot\vecg{\sigma})\,\vecg{\sigma}
-(\ve{g}_A\cdot\vecg{\sigma}-\ve{f}_A\cdot\ve{h}_A)\,\ve{h}_A,
\\
\label{ve-delta}
\vecg{\delta}_A&=&2\,\ve{g}_A-4\,(\ve{f}_A\cdot\vecg{\sigma})\,\ve{h}_A
-(\ve{g}_A\cdot\vecg{\sigma}-2\,\ve{f}_A\cdot\ve{h}_A)\,\vecg{\sigma},
\end{eqnarray}

\begin{eqnarray}\label{delta-dot-x-p-Q}
{1\over c}\,\Delta_{Q}\dot x^i_p(t)&=&
{1\over 2\,c^2}\,(1+\gamma)\,G\,\sum_A\,
\left(
\vecg{\alpha}_A\,{1\over c}\,\dot{\cal U}_A
+\vecg{\beta}_A\,{1\over c}\,\dot{\cal E}_A
+\vecg{\gamma}_A\,{1\over c}\,\dot{\cal F}_A
+\vecg{\delta}_A\,{1\over c}\,\dot{\cal V}_A
\right),
\\
\label{dot-U}
{1\over c}\,\dot{\cal U}_A&=&|\ve{d}_A|\,
{2|\ve{r}_A|-\vecg{\sigma}\cdot\ve{r}_A\over
 |\ve{r}_A|^3\,(|\ve{r}_A|-\vecg{\sigma}\cdot\ve{r}_A)^2},
\\
\label{dot-E}
{1\over c}\,\dot{\cal E}_A&=&
{|\ve{r}_A|^2-3(\vecg{\sigma}\cdot\ve{r}_A)^2\over|\ve{r}_A|^5},
\\
\label{dot-F}
{1\over c}\,\dot{\cal F}_A&=&-3\,|\ve{d}_A|\,
{\vecg{\sigma}\cdot\ve{r}_A\over|\ve{r}_A|^5},
\\
\label{dot-V}
{1\over c}\,\dot{\cal V}_A&=&-{1\over|\ve{r}_A|^3},
\end{eqnarray}

\noindent
where $\ve{h}_A=\ve{d}_A/|\ve{d}_A|$, $f^i_A=M_{ij}^A\,h^j_A$,
$g^i_A=M_{ij}^A\,\sigma^j$, and $M_{ij}^A$ is the symmetric trace-free
quadrupole moment of body A. From the point of view of the theory of
relativistic local reference systems, the multipole structure of the
gravitational field of a body is defined in the corresponding local
reference system of that body. However, for the calculation of the
gravitational light deflection due to the quadrupole field the
relativistic effects in $M^A_{ij}$ can be neglected. Matrix $M^A_{ij}$
is symmetric and trace-free and has, therefore, five independent
components which can be calculated from the second zonal harmonic
coefficient $J^A_2$ (in the case of the giant planets other
coefficients of the second order are negligible), the mass $M_A$ and
the equatorial radius $L_A$ of the planet, and the equatorial
coordinates $(\alpha^A_{\rm pole},\delta^A_{\rm pole})$ of the north
pole of its figure axis:

\begin{equation}\label{MAij-matrix}
M^A_{ij}=M_A\,L_A^2\,J_2^A\,\pmatrix{A&B&C\cr B&D&E\cr C&E&-A-D},
\end{equation}

\begin{eqnarray}
\label{A}
A&=&{1\over 3}-\cos^2\!\alpha^A_{\rm pole}\,\cos^2\!\delta^A_{\rm pole},
\\
\label{B}
B&=&-{1\over 2}\sin2\alpha^A_{\rm pole}\,\cos^2\!\delta^A_{\rm pole},
\\
\label{C}
C&=&-{1\over 2}\cos\alpha^A_{\rm pole}\,\sin2\delta^A_{\rm pole},
\\
\label{D}
D&=&{1\over 3}-\sin^2\!\alpha^A_{\rm pole}\,\cos^2\!\delta^A_{\rm pole},
\\
\label{EE}
E&=&-{1\over 2}\sin\alpha^A_{\rm pole}\,\sin2\delta^A_{\rm pole}.
\end{eqnarray}

Various post-Newtonian gravitational effects in the light propagation
(e.g. $\Delta_{pN}\ve{x}_p$ and $\Delta_{Q}\ve{x}_p$) are additive and
one can put

\begin{equation}\label{delta-x-p-additive}
\Delta\ve{x}_p=\Delta_{pN}\ve{x}_p+\Delta_{Q}\ve{x}_p.
\end{equation}

\noindent
Note that any additional post-Newtonian effects can be added in
(\ref{delta-x-p-additive}) if needed (e.g. the effects of the
rotational and/or translational motion of the gravitating bodies).


As first discussed by \citet{Klioner:1991a}, the higher-order multipole
moments of Jupiter and Saturn may produce a deflection larger than
1~\muas, provided that the source is observed very close to the
surfaces of the planets. It is easy to see that the maximal
gravitational light deflection produced by the zonal spherical
harmonics $J_n$ (all other harmonics are utterly small for the giant
planets) can be estimated as $\delta_{J_n}\sim J_n\,\delta_{pN}$. Using
modern values for $J_n$ of Jupiter and Saturn \citep[p.
342]{Weissman:et:al:1999} one can check that it is only the effects of
$J_4$ of Jupiter and Saturn which may exceed 1~\muas. Namely, the
effect of $J_4$ of Jupiter may amount to 10~\muas\ and that of Saturn
is not greater than 6~\muas. The effect of $J_n$ decreases as
$\cot^{n+1}\psi$ with the angular distance $\psi$ between the
gravitating body and the source. Therefore, the effect of $J_4$ of
Jupiter exceeds 1~\muas\ only if the angular distance $\psi$ between
the center of Jupiter and the source is smaller than 1.6 of the
apparent angular radius of Jupiter (i.e. smaller than 24 -- 39\arcsec\
depending on the mutual configuration of Jupiter and the observer).
For $J_4$ of Saturn $\psi$ should be less than 1.4 of the apparent
radius (i.e. less than 10 -- 14\arcsec\ depending on the
configuration). The influence of $J_6$ of Jupiter and Saturn also may
exceed 1~\muas\ if the real values of $J_6$ (which are known with a
large uncertainty for both planets) are larger than the values given by
\citet{Weissman:et:al:1999}. If the observations so close to Jupiter
and Saturn are to be processed, the reduction formulas for the effect
of $J_4$ can be derived from formulas given by \citet{Kopeikin:1997}.

Let us note also that, generally speaking, the standard expansion of
the gravitational potential of a body in terms of spherical functions
(with harmonic coefficients like $J_n$) converges only outside of the
sphere encompassing the body. Therefore, in general case the
calculation of the gravitational light deflection for a light ray
passing between the encompassing sphere and the surface of a body
requires some other representation of the body's gravitational
potential. As discussed above for the accuracy level of 1~\muas\ the
non-sphericity of the gravitational field must be taken into account
for the giant planets only. Since the polar radii of the giant planets
are significantly smaller than their equatorial radii, impact parameter
may be smaller than the equatorial radius of the body and, therefore,
smaller than the radius of the encompassing sphere. However, the
gravitational fields of the giant planets are close to that of a
homogeneous axisymmetrical ellipsoid $(x^2+y^2)/a^2+z^2/c^2\le1$ with
$c<a$, for which the gravitational potential can be expanded as
\citep[][p. 187]{Antonov:Timoshkova:Kholshevnikov:1988}

\begin{eqnarray}\label{ellipsoid:U}
U(r,\vartheta)
&=&{G\,M\over r}\,\left(1-\sum_{n=1}^\infty\,
\left({a\over r}\right)^{2n}\,J^{\rm el}_{2n}\,P_{2n}(\cos\vartheta)\right),
\\
\label{J_2n_ellipsoid}
J^{\rm el}_{2n}&=&(-1)^{n+1}\,{3\over (2n+1)\,(2n+3)}\,\left(1-{c^2\over a^2}\right)^n,
\end{eqnarray}

\noindent
where $M$ is the mass of the ellipsoid, $P_{n}(x)$ is the Legendre
polynomial of order $n$, and $r$ and $\vartheta$ are the radial
distance and the co-latitude, respectively. It is easy to see that
expansion (\ref{ellipsoid:U})--(\ref{J_2n_ellipsoid}) converges for
$r\ge \sqrt{a^2-c^2}$. Therefore, this expansion converges  everywhere
outside the ellipsoid (i.e. for $r\ge c$) if $c\ge a/\sqrt{2}$.
Inequality $c\ge a/\sqrt{2}$ is true for all the giant planets of the
Solar system. Moreover, since the density of the giant planets is not
constant, but gets larger for smaller $r$ one should expect that the
actual values of $J_{2n}$ are smaller than those ``predicted'' by
(\ref{J_2n_ellipsoid}). For a few first coefficients $J_{2n}$ it can be
explicitly seen: the actual values of $J_2$, $J_4$ and $J_6$ given by
\citet{Weissman:et:al:1999} are a factor 1.7--3.3 smaller than the
corresponding coefficients from (\ref{J_2n_ellipsoid}). This means that
the expansion (\ref{ellipsoid:U}) with the actual zonal harmonics
$J_{2n}$ of a giant planet converges everywhere outside of that planet.
Therefore, for the giant planets for any impact parameter of the light
ray it is sufficient to consider the standard expansion of the
gravitational potential in terms of spherical harmonics (or,
equivalently, multipole moments) and take into account the first few
coefficients as discussed above.


Coordinate velocity of the photon can be obtained by taking time
derivative of (\ref{x-p}):

\begin{equation}\label{vector-p}
\ve{p}\equiv{1\over c}\,\dot{\ve{x}}_p(t_o)=\vecg{\sigma}
+{1\over c}\,\Delta\dot{\ve{x}}_p(t_o).
\end{equation}

\noindent
The unit coordinate direction of the light propagation at the moment of
observation reads

\begin{equation}\label{vector-n-via-vector-p}
\ve{n}={\ve{p}\over|\ve{p}|}.
\end{equation}

\noindent
Eqs. (\ref{vector-p})--(\ref{vector-n-via-vector-p}) are more
convenient for numerical calculations than an analytical expansion of
$\ve{n}$ in terms of $\vecg{\sigma}$ and $\Delta\dot{\ve{x}}_p$ which
can be derived by substituting (\ref{vector-p}) into
(\ref{vector-n-via-vector-p}) and expanding in powers of $c^{-1}$.
However, the accuracy of 1 \muas\ can be attained with the simplified
first-order expansion of
(\ref{vector-p})--(\ref{vector-n-via-vector-p})

\begin{equation}\label{ve-n}
\ve{n}=\vecg{\sigma}
+{1\over c}\,\vecg{\sigma}\,\times\,(\Delta\dot{\ve{x}}_p(t_o)\,\times\,
\vecg{\sigma})
\end{equation}

\noindent
if the distance from the observer to the Sun (all other bodies play no
role here) is larger than 0.035~AU which is the case for all
currently proposed astrometrical missions. Therefore,

\begin{eqnarray}\label{ve-p-additive}
\ve{n}&=&\vecg{\sigma}+\delta\vecg{\sigma}_{pN}+\delta\vecg{\sigma}_{Q},
\\
\label{delta-pN-n}
\delta\vecg{\sigma}_{pN}&=&
{1\over c}\,\vecg{\sigma}\,\times\,(\Delta_{pN}\dot{\ve{x}}_p(t_o)\,\times\,
\vecg{\sigma}),
\\
\label{delta-Q-n}
\delta\vecg{\sigma}_{Q}&=&
{1\over c}\,\vecg{\sigma}\,\times\,(\Delta_{Q}\dot{\ve{x}}_p(t_o)\,\times\,
\vecg{\sigma}).
\end{eqnarray}

\noindent
Using (\ref{delta-dot-x-p-schwarzschild}) one gets

\begin{equation}\label{delta-pN-n-explicit}
\delta\vecg{\sigma}_{pN}=-\sum_A\,{(1+\gamma)GM_A\over c^2}\,
\,{\ve{d}_A\over |\ve{d}_A|^2}\,\left(1+\vecg{\sigma}\cdot{\ve{r}_{oA}\over|\ve{r}_{oA}|}\right).
\end{equation}

\noindent
Hence the post-Newtonian deflection angle due to the spherically
symmetric part of the gravitational field of body $A$ reads

\begin{equation}\label{delta-pN}
\delta_{pN}=
{(1+\gamma)GM_A\over c^2\,|\ve{r}_{oA}|}\,\cot {\psi_A\over2},
\end{equation}

\noindent
where $\psi_A$ is the angular distance between body $A$ and the source.
It is this formula which was used to compute the data in the third
column of Table \ref{Table-gravitational-deflection}. The effect of the
quadrupole field $\delta\vecg{\sigma}_{Q}$ can be calculated by
substituting (\ref{delta-dot-x-p-Q})--(\ref{dot-V}) and
(\ref{ve-alpha})--(\ref{ve-delta}) with (\ref{MAij-matrix})--(\ref{EE})
into (\ref{delta-Q-n}).

\begin{table}
\begin{tabular}{lrrrrrrrrr}
body    & $\delta_{pN}$&  $\psi_{\rm max}(1\muas)$
                                            &$\psi_{\rm max}(10\muas)$
                                                                & $\delta_{Q}$
                                                                          & $\psi_{\rm max}(1\muas)$
                                                                                       &$\delta_{R}$
                                                                                                & $\delta_{T^*}$
                                                                                                        & $\delta_{ppN}$
                                                                                                              & $\psi_{\rm max}(1\muas)$ \\
        &             &                     &                   &          &           &        &       &     &             \\
\hline
        &             &                     &                   &          &           &        &       &     &             \\
Sun     &$1.75\cdot10^6$&  180\degr         &180\degr           & $\sim$1  &           & 0.7    & 0.1   & 11  & 53\arcmin   \\
Mercury
        & 83          &  9\arcmin           &54\arcsec          & ---      &           & ---    & ---   & --- &             \\
Venus   & 493         &  4.5\degr           &27\arcmin          & ---      &           & ---    & ---   & --- &             \\
Earth   & 574         &  178\degr/123\degr  &154\degr/21\degr   & 0.6      &           & ---    & ---   & --- &             \\
Moon    & 26          &  9\degr/5\degr      &49\arcmin/26\arcmin& ---      &           & ---    & ---   & --- &             \\
Mars    & 116         &  25\arcmin          &2.5\arcmin         & 0.2      &           & ---    & ---   & --- &             \\
Jupiter & 16270       &  90\degr            &11.3\degr          & 240      & 152\arcsec& 0.2    & 0.8   & --- &             \\
Saturn  & 5780        &  17\degr            &1.7\degr           &  95      & 46\arcsec & ---    & 0.2   & --- &             \\
Uranus  & 2080        &  71\arcmin          &7.1\arcmin         &   8      & 4\arcsec  & ---    & ---   & --- &             \\
Neptune & 2533        &  51\arcmin          &5\arcmin           &  10      & 3\arcsec  & ---    & ---   & --- &             \\
        &             &                     &                   &          &           &        &       &     &             \\
\hline
        &             &                     &                   &          &           &        &       &     &             \\
Ganymede& 35          &  32\arcsec          & 4\arcsec          &          &           &        &       &     &             \\
Titan   & 32          &  14\arcsec          & 2\arcsec          &          &           &        &       &     &             \\
Io      & 31          &  19\arcsec          & 2\arcsec          &          &           &        &       &     &             \\
Callisto& 28          &  23\arcsec          & 3\arcsec          &          &           &        &       &     &             \\
Europe  & 19          &  11\arcsec          & 1\arcsec          &          &           &        &       &     &             \\
Triton  & 10          &  0.7\arcsec         &                   &          &           &        &       &     &             \\
Pluto   &  7          &  0.4\arcsec         &                   &          &           &        &       &     &             \\
Titania &  2.8        &  0.2\arcsec         &                   &          &           &        &       &     &             \\
Oberon  &  2.4        &  0.2\arcsec         &                   &          &           &        &       &     &             \\
Rhea    &  1.9        &  0.3\arcsec         &                   &          &           &        &       &     &             \\
Charon  &  1.7        &  0.05\arcsec        &                   &          &           &        &       &     &             \\
Iapetus &  1.6        &  0.2\arcsec         &                   &          &           &        &       &     &             \\
Ariel   &  1.4        &  0.1\arcsec         &                   &          &           &        &       &     &             \\
Ceres   &  1.2        &  0.3\arcsec         &                   &          &           &        &       &     &             \\
Dione   &  1.2        &  0.2\arcsec         &                   &          &           &        &       &     &             \\
Umbriel &  1.2        &  0.1\arcsec         &                   &          &           &        &       &     &             \\
\end{tabular}
\caption{Various gravitational effects in the light propagation in
\muas: $\delta_{pN}$ and $\delta_{ppN}$ are the post-Newtonian and
post-post-Newtonian effects due to the spherically symmetric field of
each body, $\delta_{Q}$ are the effects due to the quadrupole
gravitational fields, $\delta_{R}$ and $\delta_{T^*}$ are the effects
due to the gravitomagnetic fields caused by the rotational and
translational motions of the bodies, respectively. The estimations of
$\delta_{T^*}$ are given for the case when the coordinates of the
gravitating bodies are taken at the moment of the closest approach of
the body and the photon (see text). Symbol ``---'' means that the
corresponding effect is smaller than 0.1 \muas. Physical parameters of
the bodies are taken from \citet{Weissman:et:al:1999}. Because of the
minimal Sun avoidance angle the influence of some bodies can neglected
for certain missions (for GAIA, e.g. Mercury is too close to the Sun
and can be neglected). The angle $\psi_{\rm max}(\delta)$ is the
maximal angular distance between the body and the source at which the
corresponding effect still attains $\delta$ (the smallest possible
distance between the observer and each body is taken here; $\psi_{\rm
max}(\delta)$ is smaller for larger distances). For these estimates the
observer is supposed to be within a few million kilometers from the
Earth orbit. For the Earth and the Moon two estimates are given: for a
geostationary satellite and for a satellite at a distance of $10^6$ km
from the Earth.
\label{Table-gravitational-deflection}
}
\end{table}

\subsection{Coupling of the finite distance to the source
and the gravitational deflection: Solar system objects}
\label{Subsection-coupling-solar-system}

The next step is to convert $\vecg{\sigma}$ into the unit vector
$\ve{k}$ directed from the point of emission to the point of
observation. Let $\ve{x}_o(t_o)$ be the coordinate of the observer
(satellite) at the moment of observation $t_o$ and $\ve{x}_s(t_e)$ be
the position of the source at the moment of emission $t_e$ of the
signal which was observed at $\ve{x}_o(t_o)$. The moment of emission
$t_e$ is considered as a function of the moment of observation $t_o$
(see Section \ref{Section-proper-motion} for further discussion). Let
us denote

\begin{equation}\label{ve-R}
\ve{R}=\ve{x}_o(t_o)-\ve{x}_s(t_e),
\end{equation}

\begin{eqnarray}\label{ve-k}
\ve{k}(t_o)={\ve{R}\over |\ve{R}|},
\end{eqnarray}

\noindent
It is easy to see that vector $\ve{k}$ is
related to $\vecg{\sigma}$ as \citep{Klioner:1991a}:

\begin{eqnarray}\label{sigma-k}
\vecg{\sigma}=\ve{k}+{1\over |\ve{R}|}\,\ve{k}\,\times\,
(\Delta\ve{x}_p(t_e)\,\times\,\ve{k})+\OO4.
\end{eqnarray}

\noindent
In the case of a Solar system object (\ref{sigma-k}) can be combined with
(\ref{ve-n}) to get

\begin{eqnarray}\label{n-k}
\ve{n}&=&\ve{k}+\delta\ve{k}_{pN}+\delta\ve{k}_{Q},
\\
\label{delta-k-pN}
\delta\ve{k}_{pN}&=&
\ve{k}\,\times\,
\left(\left({1\over |\ve{R}|}\,\Delta_{pN}\ve{x}_p(t_e)+
{1\over c}\,\Delta_{pN}\dot{\ve{x}}_p(t_o)\right)
\,\times\,\ve{k}\right),
\\
\label{delta-k-Q}
\delta\ve{k}_{Q}&=&
\ve{k}\,\times\,
\left(\left({1\over |\ve{R}|}\,\Delta_{Q}\ve{x}_p(t_e)+
{1\over c}\,\Delta_{Q}\dot{\ve{x}}_p(t_o)\right)
\,\times\,\ve{k}\right).
\end{eqnarray}

\noindent
Hence,

\begin{eqnarray}\label{delta-k-pN-explicit}
\delta\ve{k}_{pN}&=&-\sum_A\,{(1+\gamma)GM_A\over c^2}\,
{\ve{R}\times(\ve{r}_{eA}\times\ve{r}_{oA})\over |\ve{R}|\,|\ve{r}_{oA}|
\,(|\ve{r}_{eA}|\,|\ve{r}_{oA}|+\ve{r}_{oA}\cdot\ve{r}_{eA})},
\\
\label{ve-r-Ae}
\ve{r}_{eA}&=&\ve{x}_s(t_e)-\ve{x}_A.
\end{eqnarray}

\noindent
The angle between vectors $\ve{k}$ and $\ve{n}$ due to $\delta\ve{k}_{pN}$
can be calculated as

\begin{equation}\label{angle-k-n}
{(1+\gamma)GM_A\over c^2\,|\ve{r}_{oA}|}\,\tan{\phi\over2},
\end{equation}

\noindent
where $\phi$ is the angle between vectors $\ve{r}_{eA}$ and
$\ve{r}_{oA}$. Note that the angle (\ref{angle-k-n}) depends on the
distance $|\ve{r}_{oA}|$ between the gravitating body and the point of
observation and on $\phi$, but does not depend on the distance
$|\ve{r}_{eA}|$ between the point of emission and the gravitating body.
For a remote source $|\ve{r}_{eA}|\to\infty$ and one has
$\phi=\pi-\psi_A$ so that (\ref{angle-k-n}) coincides with
(\ref{delta-pN}).

The effect of the quadrupole field $\delta\ve{k}_{Q}$ can be calculated by
substituting (\ref{delta-x-p-Q})--(\ref{EE}) into (\ref{delta-k-Q}).

For an object located in the Solar system a set of vectors $\ve{k}$
calculated from the observed directions $\ve{s}$ for several different
moments of time $t_o$ allows one to determine the barycentric orbit of
that object. Therefore, vector $\ve{k}$ is the final result of the
model for the Solar system objects.

\subsection{Coupling of the finite distance to the source
and the gravitational deflection: objects outside the Solar system}
\label{Subsection-coupling-stars}

The only effect in $\Delta\ve{x}_p$ which should be taken into
account here for the objects located outside of the Solar system (with
$|\ve{x}_s|>1000$ AU) is the post-Newtonian gravitational deflection
from the spherically symmetric part of the gravitational field of the
Sun. In this case from (\ref{sigma-k}) one gets

\begin{eqnarray}\label{sigma-k-Schwarzschild}
\vecg{\sigma}&=&\ve{k}-{(1+\gamma)\,GM_{\rm Sun}\over c^2 |\ve{R}|}\,
{|\ve{R}|+|\ve{r}_{oS}|-|\ve{r}_{eS}|\over|\ve{r}_{oS}\times\ve{r}_{eS}|^2}\,
\ve{R}\times(\ve{r}_{oS}\times\ve{r}_{eS})
+\OO4,
\\
\label{ve-r}
\ve{r}_{eS}&=&\ve{x}_s(t_e)-\ve{x}_{\rm Sun},
\\
\label{ve-r-0}
\ve{r}_{oS}&=&\ve{x}_o(t_o)-\ve{x}_{\rm Sun}.
\end{eqnarray}

\noindent
The angle between $\vecg{\sigma}$ and $\ve{k}$ can be calculated as

\begin{equation}
{(1+\gamma)GM_{\rm Sun}\over c^2\,|\ve{r}_{oS}|\,\sin\psi_{\rm Sun}}\,
\left(1+a-\sqrt{1-2a\cos\psi_{\rm Sun}+a^2}\right)\approx
{(1+\gamma)\,GM_{\rm Sun}\over c^2\,|\ve{R}|}\,
\left(\cot{\psi_{\rm Sun}\over2}
+{\cal O}(a)
\right),
\end{equation}

\noindent
where $a=|\ve{r}_{oS}|/|\ve{R}|$ and $\psi_{\rm Sun}$ is the angular
distance between the source and the Sun. The effect attains 8.5 \muas\
for a source situated at a distance of 1 pc and observed at the limb of
the Sun.  One can check that the angle between $\ve{k}$ and
$\vecg{\sigma}$ is larger than 1 \muas\ if $|\ve{x}_s|\le 8.5 \,{\rm
pc}$ and the source is observed within 2.3\degr\ from the Sun.  If at
least one of these conditions is violated (which is really the case for
all currently proposed astrometric missions since no observations can
be done so close to the Sun) one can put

\begin{equation}\label{sigma=k}
\vecg{\sigma}=\ve{k}.
\end{equation}

\bigskip

Let us note that the requirement to calculate the gravitational effects
with an accuracy of 1~\muas\ puts a constrain on the accuracy
of the planetary ephemerides (roughly speaking, one has to be able to
calculate the impact parameter of the light ray with respect to each
gravitating body with a sufficient accuracy). The required accuracy of
the ephemerides depends also on the minimal allowed angular distance between
the source and the gravitating body. For a grazing ray the barycentric
position of Jupiter should be known with an accuracy of 4 km and those
for the other planets with slightly lower accuracy. The barycentric
position of the Sun should be known with an accuracy of about 400 m for
a grazing ray, and with an accuracy of $\sim6000$ km for the minimal
allowed angular distance of $35\degr$ as adopted for GAIA. Note that
since the model involves relative positions of the satellite and the
gravitating bodies,
the barycentric position of the satellite must be known with at least
the same accuracy, i.e. $\sim 4$ km
(see Section 5.4 of \citet{GAIA:2000} for other
accuracy constrains).

\section{Parallax}
\label{Section-parallax}

Only the sources situated outside of the Solar system will considered
below. For the objects situated outside of the Solar system the next
step of the model is to get rid of the parallax, that is to transform
$\ve{k}$ into the unit vector $\ve{l}$ directed from the barycenter of
the Solar system to the source

\begin{equation}\label{ve-l}
\ve{l}(t)={\ve{x}_s(t_e)\over |\ve{x}_s(t_e)|}.
\end{equation}

\noindent
Note that starting from this point further parametrization of vectors
$\ve{k}$ and $\ve{l}$ formally coincides with what one could expect in
the Newtonian framework. From the formal mathematical point of view
these vectors may be considered as ``Euclidean vectors'' in 3-dimensional
coordinate space formed by the spatial coordinates of the BCRS. It is
important to understand, however, that this interpretation is only
formal and that those vectors are not Euclidean vectors in some
``underlying Euclidean physical space'', but rather integration
constants for the equations of light propagations in the BCRS. These
vectors are defined by the whole previous model of relativistic
reduction and would change if the model is changed (e.g. if another
relativistic reference system is used instead of the BCRS).

Here, the definitions of parallax, proper motion and radial velocity
compatible with general relativity at a level of 1 \muas\ (or better)
are suggested. Although the definitions are quite simple and
straightforward, their interpretation at such a high level of accuracy
is rather unusual from the point of view of classical Newtonian
astrometry. As it will be clear below parallax and proper motion are no
longer two separate effects which can be considered independently of
each other. The second-order parallaxes and proper motions as well as
the effects resulting from the interaction between these two effects
are important. Moreover, parallax, proper motion and other astrometric
parameters are coordinate-dependent parameters defined in the BCRS,
which is used as the relativistic reference system where the position
and motion of the sources are described. Therefore, these parameters have
some meaning only within a particular chosen model of relativistic
reductions. That is why the whole relativistic model of positional
observations must be considered to define these parameters and clarify
their meaning.

Let us define several parameters. The parallax of the source is
defined as

\begin{equation}\label{pi}
\pi(t_o)={1\,{\rm AU}\over |\ve{x}_s(t_e)|},
\end{equation}

\noindent
the parallactic parameter $\vecg{\Pi}$ is given by

\begin{equation}\label{ve-Pi}
\vecg{\Pi}(t_o) = \pi(t_o)\ {\ve{x}_o(t_o)\over 1\,{\rm AU}},
\end{equation}

\noindent
and finally the observed parallactic shift of the source is defined as

\begin{equation}\label{ve-pi}
\vecg{\pi}(t_o)=\ve{l}(t_o)\,\times\,
(\vecg{\Pi}(t_o)\,\times\,\ve{l}(t_o)).
\end{equation}

\noindent
With these definitions to sufficient accuracy one has

\begin{equation}\label{ve-k-l}
\ve{k}=-\ve{l}\,\left(1-{1\over 2}\,|\vecg{\pi}|^2\right)+
\vecg{\pi}\,\left(1+\ve{l}\,\cdot\,\vecg{\Pi}\right)+{\cal O}(\pi^3).
\end{equation}

\noindent
The second-order effects in (\ref{ve-k-l}) proportional to $\pi^2$ are
less 3 \muas\ if $|\ve{x}_s|\ge 1$ pc. For the accuracy of 1 \muas\ the
second-order terms can be safely neglected if $|\ve{x}_s|\ge 2$ pc.

\section{Proper motion}
\label{Section-proper-motion}

The last step of the algorithm is to provide a reasonable
parametrization of the time dependence of $\ve{l}$ and $\pi$ caused by
the motion of the source relative to the barycenter of the Solar
system.

It is commonly known that in order to convert the observed proper
motion and the observed radial velocity into true tangential and radial
velocities of the observed object additional information is required.
Since that information is not always available, the concepts of
``apparent proper motion'', ``apparent tangential velocity'' and
``apparent radial velocity'' are suggested below. These concepts
represent useful information about the observed object and should be
distinguished from the ``true tangential velocity'' and ``true radial
velocity''.  Definitions of all these concepts are discussed below.

In the present paper the following simple model for the coordinates of
the source is adopted:

\begin{equation}\label{ve-X(t)}
\ve{x}_s(t_e)=\ve{x}_s(t_e^{\ 0})+\ve{V}\,\Delta t_e
+{1\over 2}\,\ve{A}\,\Delta t_e^{\ 2}+
{\cal O}(\Delta t_e^{\ 3}),
\end{equation}

\noindent
Here, $\Delta t_e = t_e - t_e^{\ 0}$, and $\ve{V}$ and $\ve{A}$ are the
BCRS velocity and acceleration of the source evaluated at the moment of
emission $t_e^{\ 0}$ corresponding the initial epoch of observation
$t_o^{\ 0}$. This model allows one to consider single stars or
components of gravitationally bounded systems, periods of which are
much larger than the time span covered by observations. Depending on
particular properties of the source and on the time span of
observations higher-order terms in (\ref{ve-X(t)}) can also be
considered. It is also clear that in more complicated cases special
solutions for binary stars, etc. should be considered. For objects in
double or multiple systems for which it is possible to determine the
orbit, Eq. (\ref{ve-X(t)}) gives the coordinates of the center of mass
of such a system. The obvious correction should be added to the
right-hand side of (\ref{ve-X(t)}) in order to account for the orbital
motion of the object. This case will not be considered here. In this
paper we confine ourselves to Eq. (\ref{ve-X(t)}) only.

Substituting (\ref{ve-X(t)}) into the definitions of $\ve{l}$ and $\pi$ one
gets

\begin{eqnarray}\label{pi(t)-series}
\pi(t_o)&=&\pi_0
+{\dot\pi}_0\,\Delta t_e
+{1\over 2}\,{\ddot\pi}_0\,\Delta t_e^{\ 2}
+{\cal O}(\Delta t_e^{\ 3}),
\\
\label{dot-pi-0}
\dot\pi_0&=&-\pi_0\,{\ve{l}_0\cdot\ve{V}\over |\ve{x}_s(t_e^{\ 0})|},
\\
\label{ddot-pi-0}
\ddot\pi_0&=&
-\pi_0\,{\ve{l}_0\cdot\ve{A}\over |\ve{x}_s(t_e^{\ 0})|}
-\pi_0\,|\dot{\ve{l}}_0|^2
+2\,{\dot{\pi}_0^{\ 2}\over\pi_0},
\end{eqnarray}

\begin{eqnarray}\label{l(t)-series}
\ve{l}(t_o)&=&\ve{l}_0+\dot{\ve{l}}_0\,\Delta t_e
+{1\over 2}\,\ddot{\ve{l}}_0\,\Delta t_e^{\ 2}
+{\cal O}(\Delta t_e^{\ 3}),
\\
\label{dot-l-0}
\dot{\ve{l}}_0&=&
{1\over |\ve{x}_s(t_e^{\ 0})|}\,\ve{l}_0\times(\ve{V}\times\ve{l}_0),
\\
\label{ddot-l-0}
\ddot{\ve{l}}_0&=&
{1\over |\ve{x}_s(t_e^{\ 0})|}\,\ve{l}_0\times(\ve{A}\times\ve{l}_0)
-|\dot{\ve{l}}_0|^2\,\ve{l}_0
+2\,{\dot{\pi}_0\over\pi_0}\,\dot{\ve{l}}_0,
\end{eqnarray}

\noindent
where $\pi_0=\pi(t_o^{\ 0})=1\ {\rm AU}/|\ve{x}_s(t_e^{\ 0})|$ and
$\ve{l}_0=\ve{l}(t_o^{\ 0})=\ve{x}_s(t_e^{\ 0})/|\ve{x}_s(t_e^{\ 0})|$
are the parameters at the initial epoch of observation $t_o^{\ 0}$.

The signals emitted at moments $t_e^{\ 0}$ and $t_e$ are received by
the observer at moments $t_o^{\ 0}$ and $t_o$, respectively. The
corresponding moments of emission and reception are related by

\begin{equation}\label{t-T}
c\,(t_o-t_e)=|\ve{x}_o(t_o)-\ve{x}_s(t_e)|-\ve{k}(t_o)\cdot\Delta\ve{x}_p(t_o),
\end{equation}

\noindent
and by a similar equation for the moments $t_o^{\ 0}$ and $t_e^{\ 0}$.
The term proportional to $\Delta\ve{x}_p$ represents the gravitational
signal retardation (the Shapiro effect) due to the gravitational field
of the Solar system. For any source in the observed part of the
universe the Shapiro effect due to the gravitational field of the
Solar system is less than $10^{-3}$ s and can be safely neglected here.  Let
us denote $\Delta t_o = t_o - t_o^{\ 0}$ the time span of observations
corresponding the time span $\Delta t_e=t_e-t_e^{\ 0}$ of emission.
These two time intervals are related as

\begin{eqnarray}\label{DT-Dt}
\Delta t_e &=&
\left(1+{1\over c}\,\ve{l}_0\,\cdot\,\ve{V}\right)^{-1}\,
\Delta t_o
+{1\over c}\,\ve{l}_0\,\cdot\,\left(\ve{x}_o(t_o)-\ve{x}_o(t_o^{\ 0})\right)
\,\left(1+{1\over c}\,\ve{l}_0\,\cdot\,\ve{V}\right)^{-1}
\nonumber\\
&&-{1\over 2c}\,
\left(\ve{l}_0\cdot\ve{A}
+{|\ve{l}_0\times\ve{V}|^2\over |\ve{x}_s(t_e^{\ 0})|}
\right)\,\Delta t_o^{\ 2}+\dots
\end{eqnarray}

\noindent
Eq.  (\ref{DT-Dt}) results from a double Taylor expansion of the first
term on the right-hand side of (\ref{t-T}) in powers of parallax $\pi$
and $\Delta t_e$. Many terms have been neglected here since they were
estimated to produce negligible observable effects. Which terms of such
an expansion are important depends on many factors. The constrains used
here to derive (\ref{DT-Dt}) are: $\pi_0\le1\arcsec$, proper motion
$|\dot{\ve{l}}_0|\le 1\ \muas/{\rm s}\approx 32\arcsec/{\rm yr}$,
accuracy of position determination for a single observation is not
better than 1~\muas, lifetime of the mission is not longer than 5 yr,
the effects of acceleration $\ve{A}$ of the object were supposed to be
smaller than those of velocity $\ve{V}$ (i.e. $\ve{A}<2\ve{V}/\Delta t_e$).
For other constrains other terms in the expansion may become important.

The second term in (\ref{DT-Dt}) represents a quasi-periodic effect
with an amplitude $\ \sim |\ve{x}_o|/c \approx 500$~s for a satellite
located not too far from the Earth orbit. Below it will be shown that
this term gives a significant periodic term in the apparent proper
motion of the sources with sufficiently large proper motions.

It is easy to see from (\ref{pi(t)-series})--(\ref{ddot-l-0}) that the
time dependence of parallax and proper motion characterized by the
parameters $\dot\pi_0$ and $\ddot{\ve{l}}_0$ can be used to determine
the radial velocity of the source. This question has been recently
investigated in full detail and applied to Hipparcos data by
\citet{Dravins:Lindegren:Madsen:1999}. The tangential and radial
components of the barycentric velocity $\ve{V}$ of the source can be
defined by

\begin{equation}\label{tangential-real}
\ve{V}_{\rm tan}=\ve{l}_0\times(\ve{V}\times\ve{l}_0),
\end{equation}

\begin{equation}\label{radial-real}
V_{\rm rad}=\ve{l}_0\cdot\ve{V}.
\end{equation}

\noindent

Eqs. (\ref{pi(t)-series})--(\ref{ddot-l-0}) can be combined with
(\ref{DT-Dt}) to get the time dependence of $\ve{l}$ and $\pi$.
Collecting the terms linear with respect to $\Delta t_o$ we get the
definition of the apparent proper motion $\vecg{\mu}_{\rm ap}$ and the
corresponding apparent tangential velocity $\ve{V}^{\rm ap}_{\rm tan}$
as appeared in the linear term in $\ve{l}(t_o)$, and the definition of
the apparent radial velocity $V^{\rm ap}_{\rm rad}$ as appeared in the
linear term in $\pi(t_o)$:

\begin{eqnarray}\label{tangential-visual}
\ve{V}^{\rm ap}_{\rm tan}&=&\ve{l}_0\times(\ve{V}\times\ve{l}_0)\,
                      \left(1+{1\over c}\,\ve{l}_0\,\cdot\,\ve{V}\right)^{-1}
=\ve{V}_{\rm tan}\,\left(1+{1\over c}\,V_{\rm rad}\right)^{-1},
\end{eqnarray}

\begin{eqnarray}\label{proper-visual}
\vecg{\mu}_{\rm ap}&=&{1\over |\ve{x}_s(t_e^{\ 0})|}\,
                    \ve{l}_0\times(\ve{V}\times\ve{l}_0)\,
                    \left(1+{1\over c}\,\ve{l}_0\,\cdot\,\ve{V}\right)^{-1}
=\pi_0\ {\ve{V}^{\rm ap}_{\rm tan}\over 1\ {\rm AU}},
\end{eqnarray}

\begin{eqnarray}\label{radial-visual}
V^{\rm ap}_{\rm rad}&=&\ve{l}_0\cdot\ve{V}\,
                          \left(1+{1\over c}\,\ve{l}_0\,\cdot\,\ve{V}\right)^{-1}
=V_{\rm rad}\,\left(1+{1\over c}\,V_{\rm rad}\right)^{-1}.
\end{eqnarray}

\noindent
With these definitions the simplest possible models for $\pi(t_o)$ and
$\ve{l}(t_o)$ read (the higher-order terms are neglected here in order
to make this example more transparent):

\begin{eqnarray}\label{pi-Delta-t}
\pi(t_o)=\pi_0
-\pi_0^2\,{V^{\rm ap}_{\rm rad}\over 1\ {\rm AU}}\,\Delta t_o
+\dots,
\end{eqnarray}

\begin{eqnarray}\label{l-Delta-t}
\ve{l}(t_o)=\ve{l}_0 + \vecg{\mu}_{\rm ap}\,\Delta t_o
+\vecg{\mu}_{\rm ap}\,
 {1\over c}\,\biggl([\ve{x}_o(t)-\ve{x}_o(t_0)]\,\cdot\,\ve{l}_0\biggr)
+\dots.
\end{eqnarray}

Factor $\left(1+c^{-1}\,\ve{l}_0\,\cdot\,\ve{V}\right)^{-1}$ in
(\ref{tangential-visual})--(\ref{radial-visual}) has been discussed by,
e.g. \citet{Stumpff:1985} and \citet{Klioner:Kopeikin:1992}. If vectors
$\ve{l}_0$ and $\ve{V}$ are nearly antiparallel and $|\ve{V}|$ is
large, this factor may become large so that the apparent tangential and
radial velocities may even exceed light velocity. This phenomenon is
one of the well-known possible explanations of apparent superluminal
motions.

The amplitude of the third term in (\ref{l-Delta-t}) is about 170
\muas\ for the Barnard's star with its proper motion of
$10.4\arcsec$ per year
\citep[see][]{Brumberg:Klioner:Kopejkin:1990,Klioner:Kopeikin:1992}. In
general the amplitude of this effect is approximately equal to the
proper motion of the source during the time interval $\sim
|\ve{x}_o|/c$ required for the light to propagate from the observer to
the barycenter of the Solar system multiplied by cosine of the
ecliptical latitude of the source. Therefore, for a satellite not too
far from the Earth orbit where $|\ve{x}_o|/c\sim500$ s,
this effect exceeds 1~\muas\ for all stars with
the proper motion larger than $\sim63$ mas/yr. This effect is closely
related to the Roemer effect used in the 17th century to measure the
light velocity. Its potential importance for astrometry was recognized
by \citet{Schwarzschild:1894} and later discussed in detail by
\citet{Stumpff:1985}. The Roemer effect is a standard part of typical
relativistic models for pulsar timing \citep[see,
e.g.][]{Doroshenko:Kopeikin:1990}.

The apparent radial velocity $V^{\rm ap}_{\rm rad}$ can be immediately
used to calculate the true radial velocity $V_{\rm rad}$. Therefore, if
both the apparent tangential and apparent radial velocities are
determined from observations one can immediately restore the true
tangential velocity $\ve{V}_{\rm tan}$. However, even if it is not the
case the apparent velocity $\ve{V}^{\rm ap}_{\rm tan}$ can be useful by
itself. Note that the radial velocities measured by Doppler (spectral)
observations are neither true nor apparent radial velocity in our
terminology. The shift of spectral lines is affected by a number of
factors not appearing in positional observations (various gravitational
red shifts and Doppler effects; see, a detailed discussion e.g. by
\citet{Kopeikin:Ozernoy:1999}). The relativistic
effects induced by the motion of the satellite and by the gravitational
field of the Solar system in the Doppler measurements are typically of
the order of a few cm/s which is probably too small to be detectable by
space astrometry missions (for example, GAIA will measure the Doppler
shift of spectral lines with an accuracy of about 1 km/s). Therefore,
it is only the intrinsic red shift due to local physics of the object
which should be considered \citep[see, e.g.][]{Neill:1996}.

\section{Summary of the model}
\label{Section-summary}

Practical implementation of the model can be summarized as follows:

\begin{enumerate}

\item[A.] determine the orbit (the position $\ve{x}_o$ and the velocity
$\dot{\ve{x}}_o$) of the satellite with respect to the BCRS (Section
\ref{Section-motion});

\item[B.] re-parametrize the observed directions $\ve{s}$ by the
coordinate time $t$ of the BCRS (to this end Eq. (\ref{tau-s-TCB})
should be numerically integrated along the orbit of the satellite);

\item[C.] use Eqs. (\ref{s-prim})--(\ref{velocity-renormalization})
to convert the observed direction $\ve{s}$ to the source into
the unit BCRS direction $\ve{n}$ of the light ray at the point of
observation;

\item[D$_1$.] for the objects situated outside of the Solar system
use Eqs. (\ref{ve-p-additive}), (\ref{delta-pN-n-explicit}),
(\ref{delta-Q-n}) with (\ref{delta-dot-x-p-Q})--(\ref{dot-V}),
(\ref{ve-alpha})--(\ref{ve-delta}) and
(\ref{MAij-matrix})--(\ref{EE}), and (\ref{sigma=k})
to convert $\ve{n}$ into the unit BCRS direction $\ve{k}$ from the
source to the observer at the moment of observation; in order to judge
if the light deflection due to a particular gravitating body should be
considered here one can use Eq. (\ref{delta-pN}) and/or the maximal
angular distances given in Table \ref{Table-gravitational-deflection};

\item[D$_2$.] for the Solar system objects use Eqs. (\ref{n-k}),
(\ref{delta-k-pN-explicit}) and (\ref{delta-k-Q}) with
(\ref{delta-x-p-Q})--(\ref{EE}) to convert $\ve{n}$ into $\ve{k}$ (a
set of vectors $\ve{k}$ for several moments of time can be used to
determine the BCRS orbit of the object which represents the final
outcome for a Solar system body); Eqs. (\ref{delta-pN}),
(\ref{angle-k-n}) and/or the maximal angular distances given in Table
\ref{Table-gravitational-deflection} can be again used to judge if a
particular gravitating body should be taken into account here;

\item[E.] use Eqs. (\ref{ve-k-l}) with (\ref{pi})--(\ref{ve-pi}) to
take into account the parallax of the object and to convert $\ve{k}$
into the unit BCRS direction $\ve{l}$ from the barycenter of the Solar
system to the source;

\item[F.] use an appropriate model for the time dependence of $\ve{l}$ and
possibly of the parallactic parameter $\pi$ to account for the
proper motion of the source (a reasonable model for single sources is
given in Section \ref{Section-proper-motion}).

\end{enumerate}

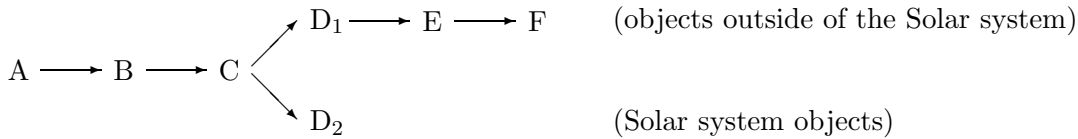
\begin{figure}
\unitlength=1.0mm
\font\rs=cmss10 scaled \magstephalf
\font\rrs=cmss10 scaled \magstep1
\begin{picture}(100.00,30.00)
\put(1,1){\makebox(0,0){\rm A}}
\put(4,1){\vector(1,0){8}}
\put(15,1){\makebox(0,0){\rm B}}
\put(18,1){\vector(1,0){8}}
\put(29,1){\makebox(0,0){\rm C}}
\put(32,1.5){\vector(1,1){6}}
\put(32,0.5){\vector(1,-1){6}}
\put(42,7.5){\makebox(0,0){\rm D$_1$}}
\put(42,-6){\makebox(0,0){\rm D$_2$}}
\put(45,7.5){\vector(1,0){8}}
\put(56,7.5){\makebox(0,0){\rm E}}
\put(59,7.5){\vector(1,0){8}}
\put(70,7.5){\makebox(0,0){\rm F}}
\put(80,7.5){\makebox(0,0)[l]{\rm (objects outside of the Solar system)}}
\put(80,-6){\makebox(0,0)[l]{\rm (Solar system objects)}}
\end{picture}
\vskip 1cm
\caption{The structure of the model for different kinds of objects (see text).
\label{Figure-structure}
}
\end{figure}

The sequence of the steps of the model is depicted on Fig.
\ref{Figure-structure}.


\section{The model for 0.1 \muas}
\label{Section-0.1-muas}

Since the technical accuracy of GAIA is expected to attain 4 \muas\ for
the stars with the magnitude of 12 mag and brighter, and the technical
accuracy of SIM is also expected to attain the level of 3-4 \muas, it
is interesting to look at the relativistic model having an accuracy of
0.1 \muas\ which guarantees that all systematic effects of the level of 1
\muas\ are properly taken into account. The relativistic effects
related to the gravitational potential of the Solar system bodies and
to the motion of the observer with respect the barycenter of the Solar
system can be modeled with the accuracy of 0.1 \muas\ within the same
scheme discussed above and summarized in the previous Section. The necessary
changes to attain the accuracy of 0.1 \muas\ are:

\begin{enumerate}

\item[Step A:] The barycentric velocity of the satellite should be known with
an accuracy of $10^{-4}$ m/s. This also requires the use of Eq. (\ref{Va-deltavi})
if the GCRS is used for orbit determination.

\item[Step D:] Some additional effects should be taken into
account in the gravitational light deflection (see also Table
\ref{Table-gravitational-deflection}):

\begin{enumerate}

\item The effect explicitly proportional to the velocity of the bodies
should be taken into account when observing within 8 angular radii of
Jupiter or 2 angular radii of Saturn. As discussed in Section
\ref{Section-gravitational-deflection} the difference between $t^{\rm
r}_{A0}$ and $t^{\rm ca}_{A0}$ still plays no role at this level of
accuracy. The explicit formulas has been published by
\citet{Klioner:1989}, \citet{Klioner:Kopeikin:1992} and
\citet{Kopeikin:Schaefer:1999}.

\item The effect of the rotational motion of Jupiter should be taken
into account when observing within 1.4 of its apparent radius
(i.e. within 42--69\arcsec\ from the Jupiter's center). The explicit
formulas for this effect have been published by \citet{Klioner:1991a}
and \citet{Klioner:Kopeikin:1992}.

\item The effect of $J_4$ of Jupiter exceeds 0.1 \muas\ within 2.5 of
its apparent radius (37--61\arcsec\ from the Jupiter's center); the
same effect for Saturn exceeds 0.1 \muas\ within 2.2 of its apparent
radius (15--22\arcsec\ from the Saturn's center). The effect of $J_6$
\citep[][p. 342]{Weissman:et:al:1999} of both Jupiter and Saturn may
exceed 0.1 \muas\ within 1.3 of the apparent radii of each planet. The
explicit formulas for the influence of $J_4$ and $J_6$ have been
published by \citet{Kopeikin:1997}.

\item All the bodies, the radius $L$ of which

\begin{equation}\label{L-rho-delta-0.1}
L\ge \left(\rho\over 1\,{\rm g/cm^3}\right)^{-1/2}\,197\,{\rm km},
\end{equation}

\noindent
where $\rho$ is the mean density of the body, can produce light
deflection of order 0.1 \muas\ and should be taken into account.
This makes it quite difficult to reduce the observations near the
ecliptic where a large number of small bodies satisfying this condition
can be expected.

\item The maximal angular distances at which the influence of a
gravitating body should be taken into account should be increased as
compared to Table \ref{Table-gravitational-deflection} to meet the goal
accuracy of 0.1 \muas. The post-post-Newtonian light deflection attains
0.1 \muas\ only for observations within 2.3\degr\ from the Sun and,
therefore, can be neglected for the planned space missions.

\item Coupling of the finite distance to the source and the gravitational
deflection for sources located outside of the Solar system (Section
\ref{Subsection-coupling-stars}) should be taken into account if
$|\ve{x}_s|\le85$~pc and the source is observed within 22.7\degr\
from the Sun. Although for the current technical specifications of the
planned space missions the effect can be neglected, it may become
important since the Sun avoidance angle of $\sim 23\degr$ is not
technically impossible.

\end{enumerate}

\item[Step E:] The second-order parallactic effects (Section \ref{Section-parallax})
should be taken into account if $|\ve{x}_s|\le5.5$ pc.

\item[Step F:] The model for the proper motion (Section \ref{Section-proper-motion})
should be correspondingly refined. In particular the third term
in (\ref{l-Delta-t}) becomes important for all the stars
with the proper motion larger than $\sim0.62$\arcsec/yr.

\end{enumerate}

Note also that additional relativistic effects due to various
gravitational fields generated outside of the Solar system (see Section
\ref{Section-beyond-the-model} below and \citet{Kopeikin:Gwinn:2000})
become even more important at the level of 0.1 \muas\ and should be
thoroughly taken into account.

\section{What is beyond the model}
\label{Section-beyond-the-model}

The relativistic model proposed above can be considered as a
``standard'' model suitable for all sources. This model allows one to
reduce the observational data with an accuracy of 1 \muas\ and restore
positions and other parameters of the objects (e.g. their velocities)
defined in the BCRS. The model properly takes into account the
gravitational field of the Solar system, but ignores a number of
possible effects which may be caused by the gravitational fields
produced outside of the Solar system. Let us review some of these
effects \citep[see also][]{Kopeikin:Gwinn:2000}.

The first additional effect to mention here is the so-called weak
microlensing which is simply an additional gravitational deflection of
the light coming from a distant source which is produced by the
gravitational field of a visible or invisible object situated near the
light path between the observed source and the observer. For
applications in high-precision astrometry one should distinguish
between microlensing events and microlensing noise. Microlensing event
is a time dependent change of source's position (and possibly its
brightness) which is large and clear enough to be identified as such.
Microlensing events can be used to determine physical properties of the
lens so that the unperturbed path of the source can be restored at the
end
\citep[e.g.][]{Hosokawa:et:al:1993,Hosokawa:et:al:1995,Hoeg:Novikov:Polnarev:1995,Belokurov:Evans:2002}.
In this sense microlensing events represent no fundamental problem for
the future astrometric missions. On the other hand, microlensing noise
comes from unidentified microlensing events (which are too weak or too
fast to be detected as such). The number of such unidentified
microlensing events will be clearly much higher than the number of
identified ones. The microlensing noise results in stochastic changes
of positions of the observed sources with unpredictable (but generally
small) amplitude and to unpredictable moments of time. Therefore,
microlensing noise can spoil the determination of positions, parallaxes
and proper motions of the objects
\citep{Zhdanov:1995,Zhdanov:Zhdanova:1995,
Hosokawa:Ohnishi:Fukushima:1997,Sazhin:Zharov:Volynkin:Kalinina:1998,
Sazhin:Zharov:Kalinina:2001,Belokurov:Evans:2002}. It is currently not
quite clear to what extent the microlensing noise produced by the
objects of the Galaxy can deteriorate the resulting catalogs of the
future astrometric missions. To clarify this question data simulations
similar to those described by
\citet{deFelice:Vecchiato:Bucciarelli:Lattanzi:Crosta:2000}, but
involving a model for microlensing with a realistic model for the
Galaxy would be of much help. The first simulation of this kind has
been performed by \citet{Belokurov:Evans:2002} who conclude that
the microlensing noise presents no problem for GAIA.
On the other hand, the accuracy level of
$\sim1$ \muas\ seems to be close to a fundamental limit of astrometric
accuracy, since at much higher accuracies the stochastic influence of
microlensing noise becomes too strong so that in too many cases measurable
relativistic deflection effects cannot be property taken into account
\citep{Sazhin:Zharov:Volynkin:Kalinina:1998,Sazhin:Zharov:Kalinina:2001}.

In edge-on binary (or multiple) systems the gravitational light
deflection due to the gravitational field of the companion may be
observable. In the case of pulsar timing observations of binary pulsars
this question has been thoroughly investigated by
\citet{Doroshenko:Kopeikin:1995}. It is clear that for companions with
stellar masses the inclination of the orbit should be very close to
90\degr\ for the effect to be observable at the level of 1 \muas. The
formulas of Section \ref{Subsection-coupling-solar-system} can be
directly used here to calculate the effect in the first post-Newtonian
approximation. If the lensing companion is a neutron star or a black
hole it is necessary to investigate the lensing effect in the strong
field regime and also consider secondary images (a detailed study of
the strong-field-regime appearance of a star orbiting a Kerr black hole
is given by \citet{Cunningham:Bardeen:1972,Cunningham:Bardeen:1973}).
Further investigation is necessary to estimate the probability to
observe a binary system for which the gravitational lensing of the
companion is important at the level of 1 \muas.

Gravitational waves can in principle produce gravitational light
deflection. Two cases should be distinguished here: 1) gravitational
waves from binary stars and other compact sources and 2) stochastic
primordial gravitational waves from the early universe. Gravitational
waves from a single compact source were shown to produce an utterly
small deflection which is hardly observable at the level of 1 \muas\
\citep{Kopeikin:Schaefer:Gwinn:Eubanks:1999}. However, the ensemble of
compact sources of gravitational waves in the Galaxy (e.g. binary
stars) may produce a larger cumulative effect which should be
thoroughly estimated (see, e.g. \citep{Kopeikin:1999} for an attempt of such an
estimation in pulsar timing measurements). The influence
of primordial gravitational waves was analyzed by
\citet{Pyne:Gwinn:Birkinshaw:Eubanks:Matsakis:1996} and
\citet{Gwinn:et:al:1997}. Although initially applied for VLBI, the
method of these authors can be directly used for optical astrometry.

Finally, cosmological effects should be accounted for to interpret the
derived parameters of the objects (e.g. the accuracy of parallaxes
$\sigma_\pi = 1$ \muas\ allows one to measure the distance to the
objects as far as $\sim1$ Mpc away from the Solar system; see, e.g.
\citet{Kristian:Sachs:1965} for a discussion of astrometric
consequences of cosmology). It may be interesting here to construct
the metric tensor of the BCRS with a cosmological solution as a background
and analyze the effects of the background cosmology in such a reference
system.

Last but not least, the observational accuracy of 1 \muas\ together
with the mission lifetime of at  least 5 years allow one to see the
apparent proper motions of quasars and other remote sources due to the
acceleration of the Solar system barycenter relative to the center of
the Galaxy \citep[][Section 1.8.10]{GAIA:2000}. These proper motions
should be of order of $\sim 4$ \muas/yr, if one assumes the Solar
system to be on a circular orbit around the Galactic center.

\acknowledgements

The author is grateful to F. Mignard, M. Soffel, S. Kopeikin and
M. Lattanzi for fruitful discussions and suggestions.


\begin{thebibliography}{}

\bibitem[Antonov, Timoshkova \& Kholshevnikov (1988)]{Antonov:Timoshkova:Kholshevnikov:1988}
Antonov, V.A., Timoshkova, E.I., \& Kholshevnikov, K.V. 1988,
Introduction to the theory of the Newtonian Potential
(Moscow: Nauka), in Russian


\bibitem[Bienayme \& Turon (2002)]{Bienayme:Turon:2002}
Bienayme, O., \& Turon, C.  2002, GAIA: A European Space Project
(Les Ulis: EDP Sciences)

\bibitem[Belokurov \& Evans (2002)]{Belokurov:Evans:2002}
Belokurov, V.A., \& Evans, N.W. 2002, \mnras, 331, 649

\bibitem[Brumberg (1986)]{Brumberg:1986}
Brumberg, V.A. 1986, in Astrometric Techniques,
ed. H.K. Eichhorn, \& R.J. Leacock (Dordrecht: Reidel), 19

\bibitem[Brumberg (1991)]{Brumberg:1991}
Brumberg, V.A. 1991, Essential Relativistic Celestial Mechanics
(Bristol: Hilger)

\bibitem[Brumberg, Klioner, \& Kopejkin (1990)]{Brumberg:Klioner:Kopejkin:1990}
Brumberg, V.A., Klioner, S.A., \& Kopejkin, S.M. 1990,
in Inertial Coordinate System on the Sky, ed. J.H. Lieske \& V.K. Abalakin
(Dordrecht: Kluwer), 229

\bibitem[Brumberg \& Kopeikin (1989a)]{Brumberg:Kopeikin:1989a}
Brumberg, V.A., \& Kopejkin, S.M. 1989a, in: Reference Frames, ed. J.
Kovalevsky, I.I. Mueller, \& B. Ko\l{}aczek (Dordrecht:Kluwer), 115

\bibitem[Brumberg \& Kopeikin (1989b)]{Brumberg:Kopeikin:1989b}
Brumberg, V.A., \& Kopejkin, S.M. 1989b, Nuovo Cimento, 103B, 63

\bibitem[Cunningham \& Bardeen (1972)]{Cunningham:Bardeen:1972}
Cunningham, C.T., \& Bardeen, J.M. 1972, \apj, 173, L137

\bibitem[Cunningham \& Bardeen (1973)]{Cunningham:Bardeen:1973}
Cunningham, C.T., \& Bardeen, J.M. 1973,
\apj, 183, 237

\bibitem[Damour, Soffel, \& Xu (1991)]{Damour:Soffel:Xu:1991}
Damour, T., Soffel, M., \& Xu, C. 1991, \prd, 43, 3273

\bibitem[Damour, Soffel, \& Xu (1992)]{Damour:Soffel:Xu:1992}
Damour, T., Soffel, M., \& Xu, C. 1992, \prd, 45, 1017

\bibitem[Damour, Soffel, \& Xu (1993)]{Damour:Soffel:Xu:1993}
Damour, T., Soffel, M., \& Xu, C. 1993,
\prd, 47, 3124

\bibitem[Damour, Soffel, \& Xu (1994)]{Damour:Soffel:Xu:1994}
    Damour, T., Soffel, M., \& Xu, C. 1994, \prd, 49, 618

\bibitem[Dravins, Lindegren, \& Madsen (1999)]{Dravins:Lindegren:Madsen:1999}
Dravins, D., Lindegren, L., \& Madsen, S. 1999, \aap, 348, 1040

\bibitem[de Felice et al. (2001)]{deFelice:Bucciarelli:Lattanzi:Vecchiato:2001}
de Felice, F., Bucciarelli, B., Lattanzi, M.G., \&  Vecchiato, A. 2001,
\aap, 373, 336

\bibitem[de Felice et al. (1998)]{deFelice:Lattanzi:Vecchiato:Bernacca:1998}
de Felice, F., Lattanzi, M.G., Vecchiato, A., \& Bernacca, P.L. 1998,
\aap, 332, 1133

\bibitem[de Felice et al. (2000)]{deFelice:Vecchiato:Bucciarelli:Lattanzi:Crosta:2000}
de Felice, F., Vecchiato, A., Bucciarelli, B., Lattanzi, M.G., \& Crosta,
M. 2000, in Towards Models and Constants for Sub-Microarcsecond Astrometry,
ed. K.J. Johnston, D.D. McCarthy, B.J. Luzum, \& G.H. Kaplan
(Washington: US Naval Observatory), 314

\bibitem[Doroshenko \& Kopeikin (1990)]{Doroshenko:Kopeikin:1990}
Doroshenko, O.V., \& Kopeikin, S.M. 1990,
\azh, 67, 986, in Russian

\bibitem[Doroshenko \& Kopeikin (1995)]{Doroshenko:Kopeikin:1995}
Doroshenko, O.V., \& Kopeikin, S.M. 1995, \mnras, 274, 1029

\bibitem[ESA (2000)]{GAIA:2000}
ESA, 2000, GAIA: Composition, Formation and Evolution of the Galaxy,
Concept and Technology Study Report, Document ESA-SCI(2000)4, European
Space Agency, Noordwijk

\bibitem[Fukushima (1995)]{Fukushima:1995}
Fukushima, T. 1995, \aap, 294, 895

\bibitem[Gwinn et al. (1997)]{Gwinn:et:al:1997}
Gwinn, C.R., Eubanks, T.M., Pyne, T., Birkinshaw, M., \& Matsakis, D.N.
1997, \apj, 485, 87

\bibitem[H\o{}g, Novikov \& Polnarev (1995)]{Hoeg:Novikov:Polnarev:1995}
H\o{}g, E., Novikov, I.D., \& Polnarev, A.G. 1995, \aap, 294, 287


\bibitem[Hosokawa et al. (1993)]{Hosokawa:et:al:1993}
Hosokawa, M., Ohnishi, K., Fukushima, T., \& Takeuti, M. 1993,
\aap, 278, L27

\bibitem[Hosokawa et al. (1995)]{Hosokawa:et:al:1995}
Hosokawa, M., Ohnishi, K., Fukushima, T., \& Takeuti, M. 1995,
in Astronomical and Astrophysical
Objectivities of Sub-Milliarcsecond Optical Astrometry,
ed. E. H\o{}g, \& P.K. Seidelmann (Dordrecht: Kluwer), 305

\bibitem[Hosokawa et al. (1997)]{Hosokawa:Ohnishi:Fukushima:1997}
Hosokawa, M., Ohnishi, K., \& Fukushima, T. 1997,
\aj, 114, 1508

\bibitem[IAU (2001)]{IAU:2001}
IAU 2001, Information Bulletin, 88 (errata in IAU Information Bulletin, 89)

\bibitem[IERS (1996)]{IERS:1996}
IERS 1996, IERS Conventions, International Earth Rotation
Service Technical Note 21, ed. D.D. McCarthy
(Paris: Observatoire de Paris)

\bibitem[Irwin \& Fukushima (1999)]{Irwin:Fukushima:1999}
Irwin, A.W. \& Fukushima, T. 1999, \aap, 348, 642

\bibitem[Klioner (1989)]{Klioner:1989}
Klioner, S.A. 1989,
Institute of Applied Astronomy, preprint No 6, in Russian

\bibitem[Klioner (1991a)]{Klioner:1991a}
Klioner, S.A. 1991a, \azh, 68, 1046, in Russian
(translated into English:  \sovast, 35, 523)

\bibitem[Klioner (1991b)]{Klioner:1991b}
Klioner, S.A. 1991b, Ph.D. thesis,
Institute of Applied Astronomy \& St.Petersburg State University,
in Russian

\bibitem[Klioner (1993)]{Klioner:1993}
Klioner, S.A. 1993, \aap, 279, 273

\bibitem[Klioner (2000)]{Klioner:2000}
Klioner, S.A. 2000,
in Towards Models and Constants for Sub-Microarcsecond Astrometry,
ed. K.J. Johnston, D.D. McCarthy, B.J. Luzum, \& G.H. Kaplan
(Washington: US Naval Observatory), 308

\bibitem[Klioner \& Kopeikin (1992)]{Klioner:Kopeikin:1992}
Klioner S.A., \& Kopeikin, S.M. 1992,
\aj, 104, 897

\bibitem[Klioner \& Soffel (1998a)]{Klioner:Soffel:1998a}
Klioner, S.A., \& Soffel, M. 1998a, \aap, 334, 1123

\bibitem[Klioner \& Soffel (1998b)]{Klioner:Soffel:1998b}
 Klioner, S.A., \& Soffel, M.H. 1998b, \prd, 58, ID 084023

\bibitem[Klioner \& Soffel (2000)]{Klioner:Soffel:2000}
Klioner, S.A., \& Soffel, M.H. 2000,
\prd, 62, ID 024019

\bibitem[Klioner \& Voinov (1993)]{Klioner:Voinov:1993}
Klioner, S.A., \& Voinov, A.V. 1993, \prd, 48, 1451

\bibitem[Kopeikin (1988)]{Kopeikin:1988}
Kopejkin, S.M. 1988, Celestial Mechanics, 44, 87


\bibitem[Kopeikin (1997)]{Kopeikin:1997}
Kopeikin, S.M. 1997, J. Math. Phys., 38, 2587

\bibitem[Kopeikin (1999)]{Kopeikin:1999}
Kopeikin, S.M. 1999, available as {\tt http://xxx.lanl.gov/abs/gr-qc/9903070}

\bibitem[Kopeikin \& Ozernoy (1999)]{Kopeikin:Ozernoy:1999}
Kopeikin, S.M., \& Ozernoy, L.M., 1999, \apj, 523, 771

\bibitem[Kopeikin et al. (1999)]{Kopeikin:Schaefer:Gwinn:Eubanks:1999}
Kopeikin, S.M., Sch\"afer, G., Gwinn, C.R., \& Eubanks, T.M. 1999,
\prd, 59, ID 084023

\bibitem[Kopeikin \& Gwinn (2000)]{Kopeikin:Gwinn:2000}
Kopeikin, S.M., \& Gwinn, C. 2000,
in Towards Models and Constants for Sub-Microarcsecond Astrometry,
ed. K.J. Johnston, D.D. McCarthy, B.J. Luzum, \& G.H. Kaplan
(Washington: US Naval Observatory), 303

\bibitem[Kopeikin \& Sch\"afer (1999)]{Kopeikin:Schaefer:1999}
Kopeikin, S.M., \& Sch\"afer, G. 1999, \prd, {\bf 60}, ID 124002

\bibitem[Kopeikin \& Mashhoon (2002)]{Kopeikin:Mashhoon:2002}
Kopeikin, S.M., \& Mashhoon, B. 2002, \prd, {\bf 65}, ID 064025

\bibitem[Kopeikin et al. (2000)]{Kopeikin:Shuygina:Vasiliev:Yagudina:Yagudin:2000}
Kopeikin, S.M., Shuygina, N., Vasiliev, M., Yagudina, E., \& Yagudin, L.
2000,
in Towards Models and Constants for Sub-Microarcsecond Astrometry,
ed. K.J. Johnston, D.D. McCarthy, B.J. Luzum, \& G.H. Kaplan
(Washington: US Naval Observatory), 320

\bibitem[Kristian \& Sachs (1965)]{Kristian:Sachs:1965}
Kristian, J., \& Sachs, R.K. 1965,
\apj, 143, 379

\bibitem[Mashhoon (1974)]{Mashhoon:1974b}
Mashhoon, B. 1974, \nat, 250, 316

\bibitem[Misner, Thorne, \& Wheeler (1973)]{Misner:Thorne:Wheeler:1973}
Misner, C.W., Thorne, K.S., Wheeler, J.A. 1973,
Gravitation (San Francisco: Freeman)

\bibitem[Neill (1996)]{Neill:1996}
Neill, R.I. 1996, \aj, 111, 2000

\bibitem[Perryman et al. (2001)]{Perryman:et:al:2001}
Perryman, M. A. C., et al. 2001, \aap, 369, 339

\bibitem[Pyne et al. (1996)]{Pyne:Gwinn:Birkinshaw:Eubanks:Matsakis:1996}
Pyne, T., Gwinn, C.R, Birkinshaw, M., Eubanks, T.M., \& Matsakis, D.N.
1996, \apj, 465, 566

\bibitem[Reasenberg et al. (1988)]{Reasenberg:et:al:1988}
Reasenberg, R.D., et al. 1988, \aj, 96, 1731

\bibitem[Sazhin et al. (1998)]{Sazhin:Zharov:Volynkin:Kalinina:1998}
Sazhin, M.V., Zharov, V.E., Volynkin, A.V., \& Kalinina, T.A. 1998,
\mnras, 300, 287

\bibitem[Sazhin et al. (2001)]{Sazhin:Zharov:Kalinina:2001}
Sazhin, M.V., Zharov, V.E., \& Kalinina, T.A. 2001, \mnras, 323, 952

\bibitem[Schwarzschild (1894)]{Schwarzschild:1894}
Schwarzschild, K. 1894, Astr. Nachr., 136, No. 3246, 81

\bibitem[Shao (1998)]{SIM:1998}
Shao, M. 1998, in Astronomical Interferometry,
ed. R.D. Reasenberg, \procspie, 3350,  536


\bibitem[Stumpff (1985)]{Stumpff:1985}
Stumpff, P. 1985, \aap, 144, 232

\bibitem[Walter et al. (1986)]{Walter:et:al:1986}
Walter, H. G., Mignard, F., Hering, R., Froeschl\'e, M., \&
Falin, J. L. 1986, Manuscripta Geodaetica, 11, 103

\bibitem[Weissman, McFadden \& Johnson (1999)]{Weissman:et:al:1999}
Weissman, P.R., McFadden, L.-A., Johnson, T.V. (eds.) 1999,
Encyclopedia of the Solar System
(San Diego: Academic Press)

\bibitem[Will (1993)]{Will:1993}
Will, C. M. 1993,
Theory and experiment in gravitational physics
(Cambridge: Cambridge University Press)

\bibitem[Zhdanov (1995)]{Zhdanov:1995}
Zhdanov, V.I. 1995, in Astronomical and Astrophysical Objectivities of
Sub-Milliarcsecond Optical Astrometry, ed. E. H\o{}g, \& P.K.
Seidelmann (Dordrecht: Kluwer), 295

\bibitem[Zhdanov \& Zhdanova (1995)]{Zhdanov:Zhdanova:1995}
Zhdanov, V.I., \& Zhdanova, V.V. 1995, \aap, 299, 321

\end{thebibliography}
\end{document}